\documentclass[aps,twocolumn,prb]{revtex4}

\usepackage{graphicx}
\usepackage{amssymb}
\usepackage{amsmath}
\usepackage{braket}
\usepackage{cleveref}

\newcommand{\R}{\mathbf{r}}

\newcommand{\A}{\text{A}}
\newcommand{\B}{\text{B}}
\newcommand{\be}{\begin{equation}}
\newcommand{\ee}{\end{equation}}
\newenvironment{acknowledgment}{{\flushleft \bf Acknowledgments:}}{}

\begin{document}

\title{Laplacian-level kinetic energy approximations based on the fourth-order gradient expansion: Global assessment and application to the subsystem formulation of density functional theory}

\author{Savio Laricchia}
\affiliation{Center for Biomolecular Nanotechnologies @UNILE,
Istituto Italiano di Tecnologia (IIT), Via Barsanti, 73010 Arnesano  (LE), Italy}
\author{Lucian A. Constantin}
\affiliation{Center for Biomolecular Nanotechnologies @UNILE,
Istituto Italiano di Tecnologia (IIT), Via Barsanti, 73010 Arnesano  (LE), Italy}
\email{lucian.constantin@iit.it}
\author{Eduardo Fabiano}
\affiliation{National Nanotechnology Laboratory (NNL), Istituto di Nanoscienze-CNR,
Via per Arnesano 16, 73100 Lecce, Italy }
\author{Fabio {Della Sala}}
\affiliation{National Nanotechnology Laboratory (NNL), Istituto di Nanoscienze-CNR,
Via per Arnesano 16, 73100 Lecce, Italy }
\affiliation{Center for Biomolecular Nanotechnologies @UNILE,
Istituto Italiano di Tecnologia (IIT), Via Barsanti, 73010 Arnesano  (LE), Italy}

\begin{abstract}
We test Laplacian-level meta-generalized gradient approximation (meta-GGA)
non-interacting kinetic energy functionals based on the fourth-order gradient expansion (GE4).
We consider several well known
Laplacian-level meta-GGAs from literature (bare GE4, modified GE4, and the MGGA functional of
Perdew and Constantin [Phys. Rev. B \textbf{75},155109 (2007)]), as well as
two newly designed Laplacian-level kinetic energy functionals (named L0.4 and L0.6). 

First, a general assessment of the different functionals is performed, testing
them for model systems (one-electron densities, Hooke's atom and different jellium systems), 
atomic and molecular kinetic energies as well as
for their behavior with respect to density-scaling transformations.
Finally, we assess, for the first time, the performance of the different functionals for 
Subsystem Density Functional Theory (DFT) calculations 
on non-covalently interacting systems.

We find that the different Laplacian-level meta-GGA kinetic functionals
may improve the description of different properties of
electronic systems but no clear overall advantage is found
over the best GGA functionals.
Concerning Subsystem DFT calculations, the
here proposed L0.4 kinetic energy functional
is competitive with state-of-the-art GGAs, whereas all other Laplacian-level
functionals fail badly.
The performance of the Laplacian-level functionals is rationalized thanks to 
a two-dimensional reduced-gradient and reduced-Laplacian decomposition of the non-additive kinetic 
energy density. 
\end{abstract}

\maketitle

\section{Introduction}
\label{sec1}
The non-interacting kinetic energy (KE) density-functional has been the
subject of intense research in electronic structure theory for almost one
century, since the introduction of the Thomas-Fermi model 
\cite{thomas26,fermi27,fermi28}.
This interest has been further motivated, and theoretically justified, by the
introduction of density functional theory (DFT) \cite{dftbook,dftbookgross}, 
which has the non-interacting KE as a main ingredient. 
In the orbital-free formulation of DFT \cite{dftbook,carterrev,chen08} the
noninteracting KE density-functional is, in fact, the main contribution to the
electronic energy which has no explicit analytical expression in terms of the
density (the other term being the exchange-correlation (XC) functional). 
Despite the effort spent on it
\cite{vW,Kirz57,Tomishina65,OP79,Murphy81,PhysRev.40.749,PhysRev.44.31,
 brack76,march77,jennings78,engel89,PhysRevA.34.4575,Scott,PhysRev.105.604,JY,
Ho,LWT92,pauli,bartolotti:4576,PhysRevA.34.4586,karasiev00,pisdies,
LLP91,QUA:QUA560390405,perdewk92,lacks94,DK87,Thak92,TW02,karasiev06,
PhysRevB.80.245120,huangcarter10,WGC98,WGC_err,PhysRevB.17.3735,PhysRevB.32.7868,
PhysRevB.53.9509,PhysRevA.54.1897,PhysRevB.57.4857,PhysRevA.57.4192,
PhysRevB.49.5220,PhysRevB.53.10589,PhysRevB.60.16350,karasiev_chap,
Acharya01121980,OL91,vitosjellium98,PhysRevB.75.155109,PhysRevB.79.115117,
alvarellos07,alvarellos08,alvarellos08b,PhysRevB.75.205122,
LC94,gotz09,apbek,apbekint,weso_chap_funct,TWESOm,PeGor,allan85}, 
however, the development of accurate
non-interacting KE functionals resulted to be an extremely difficult task.
As a consequence, orbital-free DFT is still of limited practical
utility, showing reasonable accuracy only for some solid-state
applications 
\cite{huangcarter10,WGC98,WGC_err,huang06,wangcarterof,Watson00,govind94,
zhou05,madden94,madden93,madden96}. 
Recently, the density-decomposed orbital-free DFT, that treats differently the localized and
delocalized densities, seems to bring further progress in the field 
\cite{CarterXia,PhysRevB.85.045126}.
We also acknowledge that the possibility of reaching 
chemical accuracy within orbital-free DFT calculations was proved recently, 
at least for one-dimensional systems, by calculations employing 
machine learning techniques to approximate the non-interacting 
kinetic energy (with $\sim 10^5$ parameters) \cite{Kieronml}.
This study showed however also the extreme difficulty of this problem.

On the other hand, in recent years, the interest in KE functionals was
strongly renewed by the development of density embedding methods
\cite{GoKim,senatore86,cortona,wesowarh93,wesorev,wesolowski96fhnch,huang06,hodak08,
elliot09jctc,neug10,laricchia10,Laricchia2011114,goodpaster10}, 
where a many electron system with electron
density $n(\R)$ is partitioned into two subsystems $\A$ and $\B$, such that the
total electron density is $n = n_\A + n_\B$, and the mutual interaction is
accounted for by an appropriate embedding potential. Of special relevance in
this context is the 
subsystem formulation of DFT within the Kohn-Sham formalism
\cite{wesowarh93,wesorev}. 
In the 
standard formulation of this
method the system is described by two
coupled sets of Kohn-Sham equations with constrained electron density
(KSCED \cite{wesowarh93}). Hence, 
the density is constrained to satisfy the condition $n = n_\A + n_\B$ 
by the inclusion of an external embedding potential of the form
(here the embedding potential for subsystem $\A$ is reported; a similar
expression holds for $\B$)
\begin{eqnarray}
\nonumber
&&v_{emb}^A[n_\A;n_\B](\R) =  v_{ext}^\B(\R) +  v_J[n_\B](\R) + \\
 \label{eq1}
&&\quad + \frac{\delta E_{xc}^{nadd}[n_\A;n_\B]} {\delta n_\A(\R)} + \frac{ \delta 
T_s^{nadd}[n_\A;n_\B]}{\delta n_\A(\R)}  \ ,
\end{eqnarray}
where $v_{ext}^\B$ and $v_J[n_\B](\R)$
are the external (i.e. nuclear) and the Coulomb potentials 
due to subsystem $\B$, while the non-additive
XC and kinetic energy terms are defined as
\begin{equation}
E_{xc}^{nadd}[n_\A;n_\B]=E_{xc}[n_\A+n_\B]-E_{xc}[n_\A]-E_{xc}[n_\B] \, , 
\end{equation}
\begin{equation}
T_s^{nadd}[n_\A;n_\B]=T_s[n_\A+n_\B]-T_s[n_\A]-T_s[n_\B] \label{eq:kinn} \, .
\end{equation}
Using an iterative freeze-and-thaw procedure \cite{wesorev,wesowarshel96} 
the full variational solution for the total 
system can be obtained, which is equivalent to the usual
Kohn-Sham solution, except for approximations included in the
non-additive kinetic interaction term and eventually in the 
non-additive XC contribution, if hybrid or orbital-dependent
functionals are employed in the subsystem formalism \cite{laricchia10,
Laricchia2011114,laricchia:014102,laricchia:124112}.
Henceforth, the acronym FDE will be used to refer to this
fully variational approach.

As shown by Eqs. (\ref{eq1}) and (\ref{eq:kinn}), the
accuracy of the FDE method relays on the availability
of accurate kinetic energy approximations. However, unlike
for orbital-free DFT, the FDE approach makes use not of the bare
non-interacting KE, but rather of the non-additive
KE contribution (the remaining part of the kinetic energy,
i.e. the subsystems' KE, is treated in a Kohn-Sham
fashion within the KSCED equations\cite{wesorev,wesowarshel96}). 
For non-bonded interactions, the
non-additive KE is quite small and well-behaved, so that,
in analogy to the XC energy, it can be efficiently described
by semilocal approximations. Thus, for a broad range of problems
(e.g. hydrogen bonds, dipole-dipole, and dispersion
complexes) the FDE method can reach a high performance, often below the
chemical accuracy \cite{gotz09,apbekint}.

Motivated by the practical appeal of the FDE method, in the
last years different semilocal KE approximations were
proposed to describe the non-additive KE bifunctional 
\cite{TW02,karasiev06,LC94,gotz09,apbek,apbekint,Ernzerhof200059}.
However, all these approximations are making use only of the 
simplest semilocal ingredients, i.e. the density $n$ and its gradient
$\nabla n$, being based on the generalized gradient approximation
(GGA). Comparison with the experience accumulated for the much 
widely investigated XC functional, shows nevertheless, that
the GGA level shows some inevitable limitations due to
its intrinsic simplicity, and in particular cannot
properly distinguish between different density regimes \cite{apbek,apbekint}. 

Thus, the investigation of the performance of more 
sophisticated functionals beyond the GGA level (meta-GGA functionals) 
in the context of the FDE method is of high interest.
To date, however, to our knowledge, no such study
has ever been performed. For this reason this work
has as principal goal to perform a general assessment
of some  existing \cite{Ho,PhysRevB.75.155109} 
and new Laplacian-level meta-GGA KE functionals in the context of the FDE method.
This study will be performed by first testing the general
quality of the KE functionals on a wide set of systems and 
properties. Then, direct application of the KE functionals
in FDE calculations will be considered.

At this point it is important, however, to note that
most meta-GGA XC functionals are implemented using as
additional ingredient to the GGA ones, the positive-defined
kinetic energy density $\tau=\tfrac{1}{2}\sum_i^{occ}|\nabla\phi_i|^2$,
where $\phi_i$ are the occupied Kohn-Sham orbitals. The 
Laplacian of the density instead is not used directly, 
but mimicked in the atomic core
through a function depending on $n$, $\nabla n$, and $\tau$ \cite{PKZB,tpss}.
This choice is convenient, because the positive-defined
kinetic energy density has a more regular behavior than
$\nabla^2 n$, which oscillates and diverges near the atomic nucleus.
At the same time the use of $\tau$ causes the so constructed meta-GGA functionals to have
a non-local dependence on the density (via the orbital-dependent
$\tau$). 

The focus in this paper will be on 
meta-GGA KE functionals using the Laplacian of the density
as meta-GGA ingredient.
In this way, despite some possible limitations due to the behavior
of $\nabla^2 n$, it is possible to construct a truly semilocal
KE functional, suitable to be used in the FDE formalism, and having
meta-GGA quality.
We recall in fact that $\nabla^2 n$ is an important ingredient for the 
construction of functionals and enters in the definition of
the fourth- and higher-order gradient expansion of the exact kinetic
energy \cite{Ho}.

\section{Kinetic energy functionals}
\label{sec2}
A Laplacian-level semilocal KE functional has the general form
\be
T_s[n]=\int d\R \; \tau^{TF}\;F_s(n,\nabla n,\nabla^2 n),
\label{es1}
\ee
where $\tau^{TF}=\frac{3}{10}(3\pi^{2})^{2/3}n^{5/3}$ is the 
Thomas-Fermi kinetic energy density \cite{thomas26,fermi27,fermi28}
and $F_s$ is a suitable kinetic enhancement factor.
Under a uniform scaling of the density 
($n_\lambda(\R)=\lambda^3n(\lambda\R)$, $\lambda\geq 0$), 
the exact non-interacting kinetic energy behaves as 
$T_s[n_\lambda]=\lambda^2 T_s[n]$, i.e. as the Thomas-Fermi KE.
Therefore, to have Eq. (\ref{es1}) satisfying this constraint, 
$F_s(n,\nabla n,\nabla^2 n,...)$ must be invariant under
the uniform density scaling. 
Such a goal can be achieved by considering the following
dimensionless reduced gradient and Laplacian
\begin{equation}
p=\frac{|\nabla n|^{2}}{4(3\pi^{2})^{2/3}n^{8/3}}\quad ; \quad
q=\frac{\nabla^2 n}{4(3\pi^{2})^{2/3}n^{5/3}}\ .
\end{equation}
The enhancement factor becomes therefore
\begin{equation}
F_s(n,\nabla n,\nabla^2 n)=F_s(p,q)\ .
\end{equation}

In this paper we consider the following approximations for the
kinetic enhancement factor:

(i) \textbf{Thomas-Fermi} (TF), defined as
\begin{equation}\label{es3}
F_s^{TF}=1+aq\ , 
\end{equation}
with $a$ a parameter. This is the simplest approximation and 
becomes exact for the uniform electron gas, as well for 
any region of space where the density is constant. 
The term $aq$ integrates to zero and is unimportant for the kinetic energy 
and its functional derivative. Thus, usually the parameter $a$ is set 
to zero. However, it was shown that $a=5/3$ improves 
the quality of the TF KE density \cite{YPL} (not relevant for the present work).

(ii) \textbf{Second order gradient expansion} (GE2) \cite{brack76}, defined as
\begin{equation}\label{es4}
F_s^{GE2}=1+\tfrac{5}{27}p + \tfrac{20}{9}q\ .
\end{equation}
As for the TF case, the last term in Eq. (\ref{es4}) integrates to
zero and does not contribute to the kinetic energy and potential. Therefore,
it is usually disregarded in most applications.

(iii) \textbf{Fourth-order gradient expansion} (GE4) \cite{Ho}. This
is written as 
\begin{equation}
F_s^{GE4}=F_s^{GE2} + \Delta\ ,
\label{es5}
\end{equation}
with 
\begin{equation}
\Delta=\tfrac{8}{81}q^{2}-\tfrac{1}{9}pq+\tfrac{8}{243}p^{2}\geq 0\ .
\label{es5_2}
\end{equation}
This enhancement factor is a simplified version, obtained via the 
Green's theorem integration of terms comprising higher order derivatives of
the density. It holds for finite systems under the assumption 
that $n(\mathbf{r})$ and $\nabla n(\mathbf{r})$ vanish as 
$r\rightarrow\infty$. For the full GE4 expression, see 
Refs. \onlinecite{dftbookgross,brack76}.
We note that the GE4 KE displays a serious drawback for finite systems,
as it shows the wrong behavior in the tail of the density of
a finite system. In this region in fact the density 
decays exponentially as $n\sim e^{-\alpha r}$ and the von Weizs\"{a}cker
KE is almost exact. Hence, the exact kinetic energy density behaves as
$\tau\sim\tau^{W}\sim\tau^{GE2}\sim n$. On the other hand,
we have that $\tau^{GE4}\sim n^{1/3}$, being much worse than GE2.
Moreover, the corresponding potential diverges under the same conditions.
This behavior is not surprising if we consider that 
higher-order gradient expansion terms 
are derived from small perturbations of the 
uniform electron gas, so that they contain the right physics 
for a slowly-varying density regime, but fail in rapidly-varying regions, 
such as in the tail of finite systems, or near the nucleus. 
For this reason GE4 usually worsens the atomic KE with respect to GE2. 
Similarly, the sixth-order gradient expansion (GE6) \cite{Murphy81}, that 
contains terms of order $\cal{O}$$(p^3, q^3, q^2p,...)$, has a 
kinetic energy density which even diverges in the tail of the density
of a finite system ($\tau^{GE6}\sim n^{-1/3}$). Consequently, 
the GE6 KE diverges for any finite system. 

(iv) \textbf{Modified fourth-order gradient expansion} (MGE4), 
defined by
\begin{equation}
F^{MGE4}_{s}=F^{GE4}_s\bigg/\sqrt{1+\biggl(\frac{\Delta}{1+\tfrac{5}{3}p}\biggr)^2}.
\label{FSGE4}
\end{equation}
This construction was proposed in Ref. \onlinecite{PhysRevB.75.155109} 
and recovers GE4 for a slowly-varying density. Instead,
near the nucleus and in the tail of an atom (when $|q|\rightarrow\infty$) 
$F^{MGE4}_{s}\longrightarrow 1+F_{s}^{W}$, with $F_{s}^{W}=\frac{5}{3}p$ 
the von Weizs\"{a}cker (W) enhancement factor \cite{vW}. This latter
is a much reasonable limit for rapidly-varying density regions and  
is also the correct limit for a uniform density perturbed by a 
small-amplitude short-wavelength density wave \cite{JY}.

(v) The \textbf{Laplacian-level meta-GGA}  (MGGA) \cite{PhysRevB.75.155109}, 
with the following expression
\begin{equation}
 F_{s}^{MGGA}=F_{s}^{W}+(F_{s}^{MGE4}-F_{s}^{W}) \, f_{ab}(F_{s}^{MGE4}-F_{s}^{W})\ ,
\label{pr6}
\end{equation}
where 
\begin{equation}
f_{ab}(z)=\left\{ \begin{array}{lll}
0     & z\leq 0\\
(\frac{1+e^{a/(a-z)}}{e^{a/z}+e^{a/(a-z)}})^b   & 0< z< a\\
1     & z\geq a,\\
                                    \end{array}
\right.
\label{pr4}
\end{equation}
is an 
analytical real-valued sharp-interpolating
 function, 
with $a=0.5389$ and $b=3$.
 Note that $F_s^{MGE4}-F_s^{W}$ is adimensional \cite{PhysRevB.75.155109}.
MGGA is one of the best models for the 
exact kinetic energy density, fulfilling many exact conditions, 
as the rigorous lower bound \cite{PhysRevB.75.155109,HOHO}
\begin{equation}
\tau^{W}(\mathbf{r})\leq\tau(\mathbf{r}).
\label{rlb}
\end{equation}

vi) In addition to the models listed above, in this work we consider
also a new Laplacian-level meta-GGA KE approximation
defined by the simple ansatz
\begin{equation}
F^{L\kappa}_s=1+2\kappa-\biggl(\frac{\kappa}{\displaystyle 1+\tfrac{x_1}{\kappa}}+
\frac{\kappa}{\displaystyle 1+\tfrac{x_2}{\kappa}}\biggr)\ ,
\label{es10} 
\end{equation}
with 
\begin{align}
x_1 & = \frac{5}{27}p+\Delta+\frac{\left(\frac{5}{27}p\right)^2}{\kappa} \\
x_2 &=  2\frac{\left(\frac{5}{27}p\right)\Delta}{\kappa}+\frac{\left(\frac{5}{27}p\right)^3}{\kappa^2}\ .
\label{es11}
\end{align}
%
This functional is designed to respect the following limits:
\begin{itemize}
\item[(1)] In the slowly-varying density limit ($s$ and $q\rightarrow0$)
it behaves as
\begin{equation}
F^{L\kappa}_s\rightarrow 1+\frac{5}{27}p+\Delta+\mathcal{O}(p^4,q^4,p^3q,...)\ ,
\label{es102}
\end{equation}
therefore it recovers GE4, except for the unimportant term $(20/9)q$.
Note also that Eq. (\ref{es102}) does not contain terms of 6-th order
(e.g. $(|\nabla n|)^6$), so that it recovers GE4 quite closely for a
wider range of (small) values of $s$ and $q$.
\item[(2)] In the rapidly-varying density-limit 
($s$ or $|q|\rightarrow\infty$) it behaves as
\begin{equation}
\label{esasy}
F_s^{L\kappa} \rightarrow 1 + 2\kappa \, . 
\end{equation}
Thus, it can be made to recover the 
behavior of the APBEK \cite{apbek,apbekint} or
revAPBEK \cite{apbek,apbekint} functionals in the rapidly-varying density-limit.
Consequently, we define two variants of the functional:
$\\$
$\\$
\textbf{L0.4} where $\kappa=\frac{1}{2}\kappa^{APBEK}=0.402$
$\\$
$\\$
\textbf{L0.6} where $\kappa=\frac{1}{2}\kappa^{revAPBEK}=0.623$.
\end{itemize}
 We constructed the L0.4/L0.6 functionals in order to recover the APBEK/revAPBEK limit 
because the latter functionals have been found to yield very accurate embedding energies still keeping 
good accuracy for other properties
(total and relative kinetic energies).\cite{apbek,apbekint}
On the other hand, recent works at GGA level \cite{gotz09,apbekint}
found that functionals diverging at large $s$
are very poor for the  embedding theory.
%


We remark that the MGE4, MGGA, L0.4 and L0.6 functionals  recover the GE4 limit, whereas 
other Laplacian-level meta-GGA KE functionals, e.g. those in Refs. \onlinecite{PeGor,allan85,LCPB09}, 
do not. Thus the latter functionals are not considered in this work.

\begin{figure}
\includegraphics[width=\columnwidth]{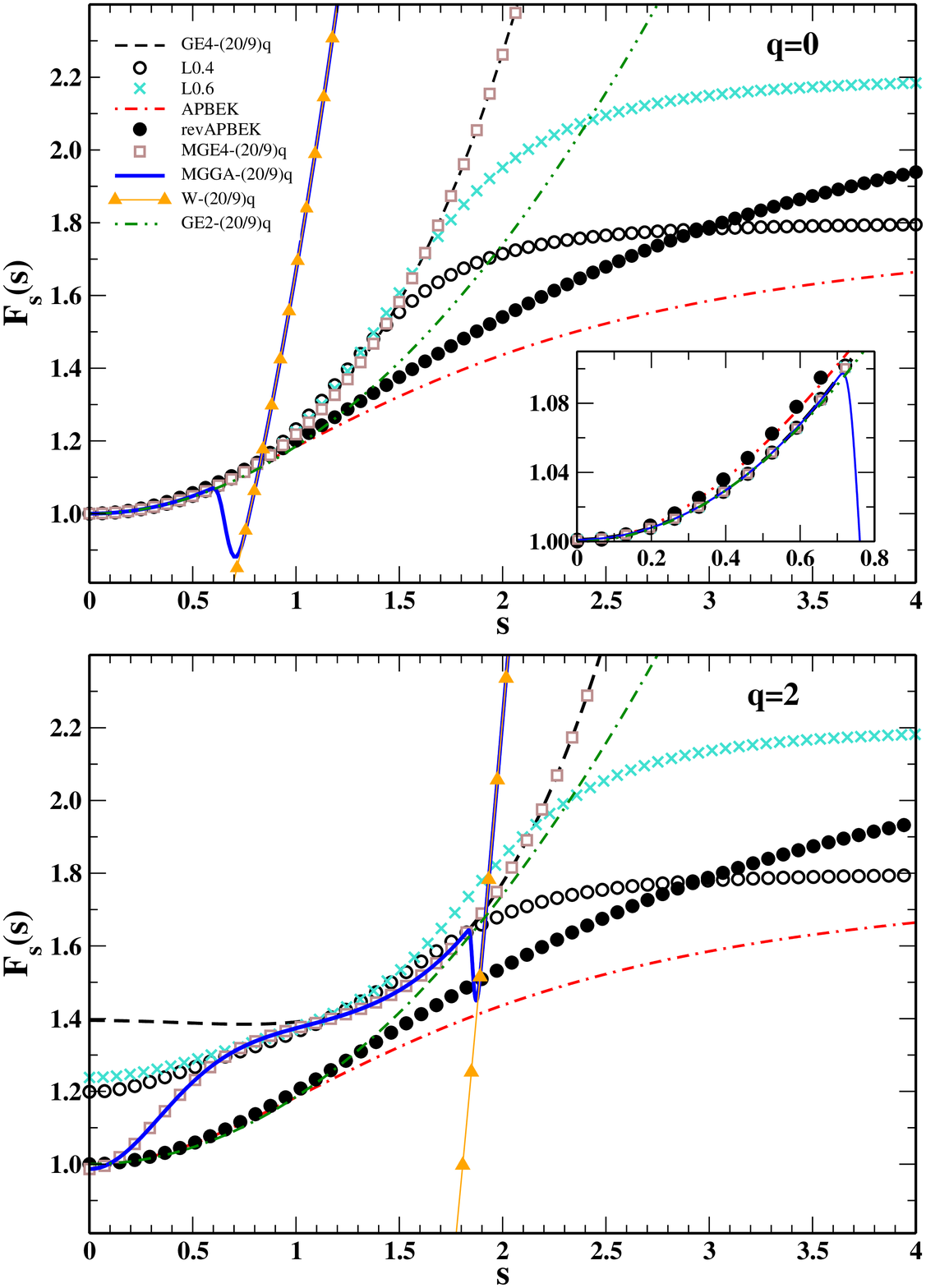}
\caption{KE enhancement factors for several functionals as functions of the 
reduced gradient $s$ ($s=\sqrt{p}$), in the case of $q=0$ (upper panel), and 
$q=2$ (lower panel). For GE4, MGE4, and MGGA we subtracted in the plot 
the term $(20/9)q$.} 
\label{f1}
\end{figure}

Figure \ref{f1} shows several enhancement factors 
as functions of the reduced gradient $s=\sqrt{p}$. The plots are reported for
two values of $q$ in the range appropriate to physical 
densities ($q=0$ and $q=2$). For 
GE2, W (von Weizs\"{a}cker)
, GE4, MGE4, and MGGA
we subtracted in the plot the term $\frac{20}{9}q$ to have a more direct
comparison with L0.4 and L0.6 that do not include such a term.
 
For $q=0$ (upper panel), all functionals 
behave similarly in the small-$s$ region ($s\lessapprox0.7$), 
recovering the modified GE2 (in case of APBEK and revAPBEK GGAs),
and the GE4 respectively (in case of Laplacian-based functionals). 
(See the inset in the upper panel of Fig. \ref{f1}).
At medium values of $s$ ($0.7\lessapprox s\lessapprox1.7$)
MGE4, L0.4, and L0.6 still recover, by construction, the GE4 behavior.
On the other hand, MGGA shows an
unphysical strong oscillation, due to the sharp interpolation function (Eq. (13)).
Finally, at large values of $s$, GE4, MGE4, and MGGA diverge, whereas 
the L0.4 and L0.6 functionals show a saturation towards the
APBEK and revAPBEK limits. Note however that these limits are only
reached at very large values of the reduced gradient.
Moreover, all the functionals, but MGGA, fail to respect the exact constraint $F_s \geq F_s^{W}$.

For $q=2$ (lower panel), a moderately-varying density regime is considered.
The picture is not much changing for large $s$ values,
but it is drastically modified for small values of the 
reduced gradient. 
Note however, that
for the GGA functionals (APBEK and revAPBEK) we have of course
the same plot as for the previous case.
In this case, at medium values of $s$ all Laplacian-level meta-GGA functionals agree rather 
well, except for the unphysical oscillations displayed by MGGA.
For small $s$-values ($s\rightarrow 0$) instead different behaviors are observed:
MGE4 and MGGA have the same trend and move towards Thomas-Fermi;
GE4 tends to $1+(8/81)q^2\approx1.395$ displaying its well known
divergent behavior in this regime; and finally, the L0.4 and L0.6
functionals tend to
\begin{equation}
F_s^{L\kappa}\rightarrow 1 + 2\kappa 
-\left(\frac{\kappa}{1+\frac{8}{81}\frac{q^2}{\kappa}}+\kappa\right)\ ,
\label{eq100}
\end{equation}
that is $F_s^{L0.4}\rightarrow1.20$ and $F_s^{L0.6}\rightarrow1.24$.

%
\begin{figure}
\includegraphics[width=\columnwidth]{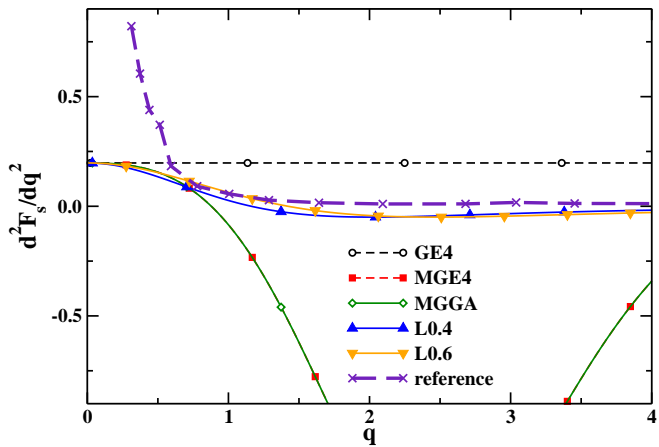}
\caption{$\partial^2 F_s(p=0,q)/\partial q^2$ versus $q$ for 
several enhancement factors.}
\label{f1b}
\end{figure}
%
The density regime characterized by $s=0$ and $q$ finite, is important 
in the middle of molecular bonds. In order to understand better this 
density regime, we report in Fig. \ref{f1b},
$\partial^2 F_s(s=0,q)/\partial q^2$ versus $q$ for
several enhancement factors. Note that $\partial^2 F_s(s,q)/\partial q^2$ is an 
intrinsic quantity of the enhancement 
factor, being independent on the linear $\frac{20}{9} q$ term.
As reference, we consider the exact Kohn-Sham KE enhancement factor derivative 
($\partial^2 F_s^{exact}(s=0,q)/\partial 
q^2$) (computed numerically) in the middle of the bond 
between two interacting jellium slabs of $r_s=3$, and thickness 
$2\lambda_F$ each (with $\lambda_F=2\pi/k_F$ being the 
Fermi wavelength). 
When the jellium slabs are close to each other ($z\ll\lambda_F$), 
$q$ is small (and positive) at the bond, and 
when the jellium slabs are far to each other ($z\geq\lambda_F$), $q$ is large (and positive). 
While MGE4 and MGGA show strong oscillations, the L$\kappa$ functionals perform
remarkably smooth, and close to the reference. Note that in case $s=0$ and $q\rightarrow 0$, 
the reference curve contains higher order terms that are out of reach for GE4-based functionals. 

\section{Density-scaling behavior}
\label{sec46}
Consider the family of density scalings \cite{scaling}
\begin{equation}
n_\lambda(\R) = \lambda^{3\beta+1}n(\lambda^\beta\R), \;\;\;\; \lambda > 0 ,
\label{es61}
\end{equation}
where $\beta$ is a parameter, which is changing not only the
external potential associated with the density $n$ (as in the uniform density scaling),
but also the particle number ($N\to\lambda N$).
Under this scaling, the KE gradient expansion terms behave as \cite{scaling}
\begin{subequations}
\begin{align}\label{es62a}
T_s^{TF}[n_\lambda] & =  \lambda^{2\beta+5/3}T_s^{TF}[n],  \\
T_s^{(2)}[n_\lambda] & = \tfrac{1}{9}T_s^{W}[n_\lambda] =
\tfrac{1}{9} \lambda^{2\beta+1}T_s^{W}[n], \label{es62b} \\
T_s^{(4)}[n_\lambda] & =  \lambda^{2\beta+1/3}T_s^{(4)}[n]\ , \label{es62c}
\end{align}
\end{subequations}

where $T_s^{(2)}=\int d\R\;\tau^{TF}[\frac{5}{27}p+\frac{20}{9}q]$ and
$T_s^{(4)}=\int d\R\;\tau^{TF}\Delta(p,q)$ are the second- and fourth-order terms of the
KE gradient expansion, respectively. On the other hand, the reduced gradient
$s$ and the reduced Laplacian $q$ scale according to
$s_\lambda(\R)=\lambda^{-1/3}s(\lambda^\beta\R)$ and
$q_\lambda(\R)=\lambda^{-2/3}q(\lambda^\beta\R)$, respectively. Thus, the reduced
gradient and Laplacian are independent on $\beta$ and the slowly varying
density limit ($s,q \to 0$) is reached whenever $\lambda\to\infty$. For
$\lambda\ll 1$, instead, a rapidly varying density regime is always set up.

For different values of the parameter $\beta$, Eq. (\ref{es61})
spans an impressive number
of physical properties. Some of them are analyzed in next subsections.

\subsection{Thomas-Fermi scaling}
For $\beta=\frac{1}{3}$ the Thomas-Fermi scaling is obtained. In the limit
$\lambda\rightarrow\infty$ any system resembles the features of the
Thomas-Fermi density \cite{Lieb1}. 
The Thomas-Fermi scaling is the basis for the
asymptotic expansion of the kinetic energy \cite{LCPB09,elliot09,PhysRevLett.100.256406}
\begin{equation}
T_s=c_0\lambda^{7/3}+c_1\lambda^2+c_2\lambda^{5/3}+... \;,
\label{eee}
\end{equation}
with the coefficients $c_0=0.768745$, $c_1=-0.5$, and
$c_2=0.2699$, fixed from the the semiclassical theory of the
neutral atom.
The expansion of Eq. (\ref{eee}) is very accurate
even for real atoms with errors of the order of 0.5\%-0.2\%.

Using the method proposed in Ref. \onlinecite{LCPB09}, we extracted
the coefficients $c_1$ and $c_2$ for all the functionals considered in this
work. We assumed that, because all the functionals recover TF for a constant density
, they have the exact $c_0$ coefficient. 

The deviations
\begin{equation}
\Delta c_i=1000(c_i^{approx}-c_i^{exact}) \qquad i=1,2 \, ,
\label{eec}
\end{equation}
are reported in Table \ref{tab2} for each functional.
%
\begin{table*}
\begin{center}
\caption{\label{tab2} Error statistics for different density scaling tests. 
The best Laplacian-level meta-GGA result in each line is
highlighted in bold face; the worst Laplacian-level meta-GGA result for each test is underlined.}
\begin{tabular}{lrrrrrrrrrr}
\hline\hline
      &      & \multicolumn{3}{c}{GGAs} & $\;\;$ & \multicolumn{5}{c}{meta-GGAs} \\
\cline{3-5}\cline{7-11}
      &  TF  &  GE2  &APBEK  &  revAPBEK  & &  GE4  &  MGE4  &  MGGA  &  L0.4  &  L0.6 \\
\hline
\multicolumn{11}{c}{\textbf{Thomas-Fermi scaling ($\beta=1/3$)}} \\
$\Delta c_1$ & -160.87 & -36.25 & -3.77 & -2.20 & & -20.39 & -17.28 & \textbf{-8.90} & \underline{-23.40} & -22.39 \\
$\Delta c_2$ &  115.50 &  66.13 & 17.14 & 27.00 & &  73.25 &  72.84 & \textbf{47.53} &  \underline{79.04} &  74.75 \\
             &         &        &       &       & &        &        &                &                    &  \\
\multicolumn{11}{c}{\textbf{Homogeneous scaling ($\beta=0$)}} \\
$\Delta s_{eff}$ & 0.67 & 0.55 & 0.59 & 0.58 & & 0.47 & 0.49 & \textbf{0.02} & \underline{0.54} & 0.53 \\
             &         &        &       &       & &        &        &                &                    &  \\
\multicolumn{11}{c}{\textbf{Fractional scaling ($\beta=1$)}} \\
$\Delta s_{eff}$ & 0.67 & 0.47 & 0.50 & 0.50 & & 0.38 & 0.42 & \textbf{0.02} & \underline{0.45} & 0.44 \\
$\Delta$        & 12.26 & 34.71 & 33.84 & 35.89 & & 38.05 & \underline{38.89} & \textbf{7.77} & 37.15 & 37.15 \\
\hline\hline
\end{tabular}
\end{center}
\end{table*}
%
As it might be expected, the best results are found for
revAPBEK and APBEK which were constructed from the semiclassical
theory of the neutral atom. 
The Laplacian-level meta-GGAs provide a good performance,
improving over GE2 for $c_1$, which is the leading term 
of quantum effects, beyond the Thomas-Fermi theory.

\subsection{Uniform-electron-gas scaling}
The uniform-electron-gas scaling is obtained for $\beta=-\frac{1}{3}$ and
$\lambda\rightarrow\infty$. Under these conditions, in fact, the
density becomes very slowly varying, being almost constant over the space.
Hence, the gradient expansion
\begin{equation}
T_s^{exact}[n_\lambda]=T_s^{TF}[n_\lambda]+T_s^{(2)}[n_\lambda]+
T_s^{(4)}[n_\lambda]+T_s^{(6)}[n_\lambda]+... \;,
\label{eee1}
\end{equation}
which was derived from small perturbations of
the uniform electron gas, is (almost) exact in this limit.
As a consequence all the Laplacian-level meta-GGA kinetic functionals,
which recover GE4 in the slowly varying density,
become very accurate under the uniform-electron-gas scaling.

\subsection{Homogeneous scaling}
Setting $\beta=0$ and considering the limit $\lambda\rightarrow 0$
the homogeneous scaling is realized
\cite{scaling,PhysRevA.59.2670,nagy:044105,doi:10.1021/cr200107z,PhysRevA.55.1792,Parr1997164}.
This scaling is a valuable tool in DFT and was used to study
the static correlation \cite{cohen:121104,Cohen08082008}, and the delocalization error \cite{Cohen08082008}, 
as well as to construct
kinetic energy functionals \cite{doi:10.1021/ct400129d}.

Following Ref. \onlinecite{scaling} we study the
ability of different functionals to satisfy the
homogeneous scaling by considering the hydrogen density
$n_H$ and the associated effective scaling order
\begin{equation}
s_{eff} = \int_0^1\frac{\ln\left(\left|T_s[n_H]\right|\right)-\ln\left(\left|
T_s[n_{H\lambda}]\right|\right)}{\ln(\lambda)}d\lambda\ .
\label{e68}
\end{equation}
This provides a measure for the scaling behavior of a generic KE functional.
The exact result is $s_{eff}^{exact}=1$ (see eq. \ref{es62b}).

The errors on $s_{eff}$ for different functionals are
reported in Table \ref{tab2}.
All the Laplacian-level meta-GGA functionals perform slightly better
than GGA ones, showing that the inclusion
of the Laplacian dependence can help to improve
the physical behavior of the functional.
Notably, the MGGA functional outstands over the other meta-GGAs, 
scaling almost perfectly under the homogeneous scaling.
This finding supports the conclusion that
this sophisticated approximation can capture very well the physics of
the von Weizs\"{a}cker functional.

\subsection{Fractional scaling}
The fractional scaling is defined by $\beta=1$ and the limit
$\lambda\rightarrow0$. It describes the physics of systems
with a fractional number of electrons \cite{scaling}.
In particular, the fractional scaling is related to the disintegration
of the hydrogen atom into fragments with fractional nuclear
and electronic charge, which is a model for atomization processes in molecules
\cite{scaling}.

In analogy with Ref. \onlinecite{scaling} we define the
kinetic disintegration energy of the hydrogen atom into
fragments with partial charge $\lambda$ and $1-\lambda$ as
\begin{equation}
M(\lambda)=T_s(1)-T_s(\lambda)-T_s(1-\lambda),
\label{e64}
\end{equation}
where $T_s(1)$ is the kinetic energy of the hydrogen atom
and $T_s(\lambda)$ the KE of the neutral fragment with charge $\lambda$.
The disintegration error is then defined as
\begin{equation}
\Delta M = \int_0^1\left[M^{DFT}(\lambda)-M^{exact}(\lambda)\right]d\lambda.
\label{e65}
\end{equation}

In Table \ref{tab2} we report, for all the functionals
the errors on the effective scaling order (Eq. (\ref{e68})), with respect to
the exact result $s_{eff}=3$, 
and the disintegration error of Eq. (\ref{e65}).
The results show that the effective scaling order follows
a similar trend as for the homogeneous scaling.
The inspection of
the values of $\Delta M$ provides a couple of additional considerations, that 
show the importance of fractional scaling:
(1) The MGGA functional works very well for both $s_{eff}$ and the
disintegration problem, showing that this functional can
incorporate most of the von Weizs\"{a}cker \cite{vW} physics
without error compensations; (2) The TF functional displays the
worst $s_{eff}$ value, but the smallest error for the
disintegration problem (except for MGGA), showing 
an important error cancellation.

\section{Computational details}
\label{sec3}

\subsection{Kinetic energies}
To test the different functionals we assessed their ability to
compute the KE of different systems:
\begin{itemize}
\item \textbf{Model one- and two-electron systems}. These include
the one-electron Gaussian, hydrogen, and cuspless-hydrogen densities, as
well as the Hooke's atom at various values of the harmonic potential.
For these systems the reference values were computed via the 
von Weizs\"{a}cker formula. In all calculations exact
densities were employed.
\item \textbf{Jellium systems}. 
We considered:
{\bf (i)} a series of Na jellium clusters ($r_s=3.93$) with magic
electron number 2, 8, 18, 20, 34, 40, 58, 92, and 106, used also in 
Refs. \onlinecite{apbekint,PhysRevB.75.155109,PhysRevB.79.115117};
{\bf (ii)} jellium surfaces with
bulk parameter $r_s$=2, 4, and 6 into the liquid drop model (LDM) 
(as in Refs. \onlinecite{apbekint,PhysRevB.75.155109,PhysRevB.79.115117});
and {\bf (iii)} two interacting jellium slabs at different distances.
Each jellium slab has $r_s=3$ and a thickness of $2\lambda_F$.

All the calculations were performed using the orbitals and densities
resulting from numerical Kohn-Sham calculations within the
local density approximation \cite{KS} for the exchange-correlation
functional ({\bf (ii)} and {\bf (iii)}), and 
the exact exchange
functional ({\bf (i)}).

\item \textbf{Atoms and ions} We tested 
the benchmark set of atoms and ions used in 
Refs. \onlinecite{apbekint,PhysRevB.75.155109,PhysRevB.79.115117}.
All calculations employed analytic Hartree-Fock orbitals and 
densities \cite{CR74}. We also calculated the ionization KE of 
noble atoms (until Uuo), using accurate exact-exchange Kohn-Sham
densities and orbitals.
\item \textbf{Molecules}. We considered the set of molecules including
H$_2$, HF, H$_2$O, CH$_4$, NH$_3$, CO, F$_2$, HCN, N$_2$, CN, NO, and O$_2$.
This set was already used in 
Refs. \onlinecite{PhysRevB.75.155109,PhysRevB.79.115117,IEMS01}.
The noninteracting kinetic energies of test molecules
were calculated using the PROAIMV code \cite{KBT82}.
The required Kohn-Sham orbitals were obtained by Kohn-Sham calculations
performed with the uncontracted 6-311+G(3df,2p) basis set, the
Becke 1988 exchange functional \cite{BEC88}, and Perdew-Wang
correlation functional \cite{PW91}.
\end{itemize}

\subsection{FDE calculations}
\label{sec:fdecalc}

The FDE calculations were performed using the \texttt{FDE} script as
implemented in the \texttt{TURBOMOLE} program package 
\cite{TURBOMOLE}. 
Details about our KSCED implementation in \texttt{TURBOMOLE} are discussed in 
Refs. \onlinecite{laricchia10,apbekint}. 
The implementation of Laplacian-level meta-GGA functionals
is briefly discussed in the appendix.
In all calculations, 
the PBE \cite{pbe} XC functional and def2-TZVPPD monomolecular basis set
\cite{def2tzvpp,furchepol} were used. 
The use of a monomolecular approach was needed to guarantee good
convergence for calculations using the MGE4 and MGGA KE functionals.
These functionals show in fact marked oscillations in the KE potential
in the tail of the density. Therefore, they can give rise to important numerical
problems in a supermolecular basis-set calculation.
The quality of our results was tested by comparing monomolecular
and supermolecular results for the GGA KE functionals as well
checking the convergence with increasingly large basis sets.
We found that the def2-TZVPPD basis set, adding diffuse basis functions to the 
def2-TZVPP \cite{def2tzvpp} basis, provides finally a reliable
description of all the systems considered in this paper.

The FDE calculations were performed on the following test systems,
characterized by different interaction characters:
He-Ne, He-Ar, Ne-Ne, Ne-Ar, CH$_4$-Ne, C$_6$H$_6$-Ne, CH$_4$-CH$_4$
(weak interaction); H$_2$S-H$_2$S, HCl-HCl, H$_2$S-HCl, CH$_3$Cl-HCl,
CH$_3$SH-HCN, CH$_3$SH-HCl (dipole-dipole interaction);
NH$_3$-NH$_3$, HCl-HCl, H$_2$O-H$_2$O, NH$_3$-H$_2$O, HF-NCH,
(HCONH$_2$)$_2$, (HCOOH)$_2$ (hydrogen bond).
The geometry of the complexes was taken from 
Refs. \onlinecite{truhlar05a,truhlar05nb,wesolowski96fhnch}.
The subsystems A and B are the monomer units.

The error on the total embedding energy was computed as 
\cite{laricchia10,laricchia:014102}
\begin{equation}
\Delta E=E^{FDE}[n_\A;n_\B]-E^{KS}[n^{KS}],
\label{deltaee}
\end{equation}
where $E^{FDE}[n_\A;n_\B]$ is the FDE total energy obtained from the
embedded subsystem densities $n_\A$ and $n_\B$, whereas $E^{KS}$ is the
conventional Kohn-Sham total energy corresponding to the ground state density
$n^{KS}$.
The performance of the different approaches was
evaluated, within each group of molecules, by computing the mean absolute
error (MAE). 

The errors on the embedding densities were studied
by considering the deformation density
\begin{equation}
\Delta n(\mathbf{r}) = n_\A(\mathbf{r}) + n_\B(\mathbf{r}) - n^{KS}(\mathbf{r})\ .
\end{equation}
Some plots for different systems were realized by 
representing the plane-averaged deformation density
\begin{equation}
\langle \Delta n\rangle_{xy}(z) = \int\int \left|\Delta n(x,y,z)\right| dxdy\ , 
\end{equation}
where we used Cartesian coordinates explicitly and the $z$ direction is along the intermolecular axis.
Finally, a quantitative measurement of the absolute error associated with a given 
embedding density was obtained by computing the embedding density error
\begin{equation}\label{xi}
\xi=\frac{1000}{N}\int\left|
\Delta n(\R) \right|\,d \mathbf{r},
\end{equation}
with $N$ the number of electrons. In the evaluation of 
$\xi$, only
valence electron densities were considered;
Core densities are in fact much higher than valence ones and would
largely dominate the calculation of $\xi$.
On the other hand, core densities are not very important for the 
determination of chemical and physical properties of the interaction 
between the subsystems, which are of interest here.

\section{Kinetic energies of model systems}
\label{sec4}
In this section we present the KE results of 
one-electron densities, the Hooke's atom, jellium surfaces, jellium
clusters, interacting jellium slabs, atomic systems, and molecules.  
%
\begin{table*}[tb]
\begin{center}
\caption{\label{tab1} 
Mean absolute relative errors (MARE) in \% ,
for different tests, and absolute relative errors for one-electron densities (G,H, and C).
The best Laplacian-level meta-GGA result in each line is highlighted in bold face; the worst
Laplacian-level meta-GGA result for each test is underlined. The last line reports the MARE relative to the 
GE2 performance (see text for details). 
The star symbol stands for a Laplacian-level meta-GGA KE functional that is better than 
the best GGA.}
\begin{small}
\begin{tabular}{lrrrrrrrrrr}
\hline\hline
      &      & \multicolumn{3}{c}{GGAs} & $\;\;$ & \multicolumn{5}{c}{meta-GGAs} \\
\cline{3-5}\cline{7-11} 
      &  TF  & 	GE2  &APBEK  &  revAPBEK  & &  GE4  &  MGE4  &  MGGA  &  L0.4  &  L0.6 \\
\hline
\multicolumn{11}{c}{\textbf{Total kinetic energies}} \\
H                &  8.2 &  2.9 & 2.2 & 3.1 & &  \underline{5.8} & 5.5 & \textbf{2.5} & 4.4 & 4.5 \\
G                & 10.1 &  1.0 & 1.8 & 0.8 & & \underline{15.3} & 4.2 & 3.7 & *\textbf{0.7} & 1.3 \\
C                &  5.3 &  5.8 & 4.6 & 5.6 & &  8.9 & 7.9 & *\textbf{2.1} & 6.8 & \underline{9.6} \\
Hooke's atom KE  &  25.7 & 14.6 & 18.8 & 17.8 &&  14.6 & *\textbf{10.3} & *14.3 & \underline{15.5} & *14.5 \\
Jell. clust. KE  &   4.4 &  1.0 & 1.0 & 0.8 & &  1.9 &  1.2 & \underline{2.5} & \textbf{0.9} & 1.0 \\
Jell. slabs KE      & 1.89 & 0.57 & 0.55 & 0.46 & & *\textbf{0.21} & *0.36 & *0.36 & *\underline{0.42} & *0.39 \\
Atoms' KE        & 8.4 & 1.1 & 0.8 & 1.2 & & \underline{2.5} &  2.4 & \textbf{1.4} & 1.9 & 2.0 \\
Molecules' KE    & 9.7 & 0.9 & 0.5 & 0.4 & & 1.0 & 1.2 & \underline{1.4} & \textbf{0.8} & \textbf{0.8} \\
\hline
MARE(GE2)          & 5.42 & 1 & 0.94 & 0.88 & & \underline{2.16} & 1.42 & 1.43 & \textbf{1.07} & 1.15 \\
\hline
\multicolumn{11}{c}{\textbf{Kinetic energy differences}} \\
Jell. clust. DKE &  17.7 & 27.2 & 18.9 & 23.1 &&  \underline{50.2} & 39.0 & *\textbf{17.9} & 29.3 & 31.5 \\
Jell. surf. LDM  & 8.1 & 3.3 & 3.9 & 3.6 & & *\textbf{1.7} & *1.9 & *2.5 & *\underline{2.8} & *\underline{2.8} \\
Jell. slabs DKE  & 17.28 & 5.02 & 4.17 & 3.45 & & *\textbf{1.31} & \underline{3.53} & \underline{3.53} & *2.85 & *2.65 \\
Atoms' IKE       & 49.5 & 42.2 & 45.8 & 44.4 & & *\textbf{39.2} & *40.7 & \underline{51.8} & *41.3 & *41.4  \\
Molecules' AKE   & 106.0 & 184.0 & 142.0 & 155.0 & & \underline{222.0} & 207.0 & *\textbf{108.0} & 216.0 & 217.0  \\
\hline
MARE(GE2)         & 1.66 & 1.00 & 0.91 & 0.90 & & 0.95 & \underline{0.96} & *\textbf{0.79} & 0.93 & 0.94 \\
\hline
\hline\hline
\end{tabular}
\end{small}
\end{center}
\end{table*}
%
All results are summarized in Table \ref{tab1}, where we report 
mean absolute relative errors (MARE) for each test.
In addition, the last line shows the average performance relative to the GE2 method, defined as
\begin{equation}
\mathrm{MARE(GE2)} = \frac{1}{N}\sum_{i=1}^N\frac{\mathrm{MARE}_i}{\mathrm{MARE_{GE2}}}\ ,
\end{equation}
where the sum runs over all the $N$ tests and MARE$_i$ is the MARE of the $i$-th test.

\subsection{One-electron densities}
\label{sec41}
We tested the different KE functionals on three model one-electron densities,
namely, the hydrogen (H), the Gaussian (G), and the cuspless densities.
These are defined as
\begin{equation}
n_H(r)=\frac{e^{-2r}}{\pi},\;\;
n_G(r)=\frac{e^{-r^2}}{\pi^{3/2}},\;\;n_C(r)=\frac{(1+r)e^{-r}}{32\pi}.
\label{es41}  
\end{equation}
They were used in the construction of several XC functionals \cite{tpss,zeta,vzeta}, 
being simple models for simple iso-orbital regions. 

For these model densities,
the von Weizs\"{a}cker \cite{vW} functional is exact and behaves as
$\tau^{W}\sim n$. A similar behavior is found
for GE2 and all the GGAs, which therefore perform rather well for
this problem, with errors below 6\%.
On the other hand, for GE4, in iso-orbital regions, we have
$\tau^{GE4}\sim n^{1/3}$. For this reason GE4
performs significantly worse than GE2 in all cases (errors up to 15\%)
(note that it is in general also worse than TF, that has
$\tau^{TF}\sim n^{5/3}$).
Finally, the correct behavior is restored for
the other Laplacian-level meta-GGAs, which thus describe these
one-electron densities reasonably well. 
In particular, in two cases, Laplacian-level meta-GGA functionals are more accurate than the best GGA functional:
MGGA is the most accurate 
for the delocalized C density, and
the L0.4 functional is the most accurate for the Gaussian density.

\subsection{Hooke's atom}
\label{sec42}
The Hooke's atom consists of two interacting electrons
in an isotropic harmonic potential of frequency $\omega$.
At small values of $\omega$, the electrons are strongly correlated.
At large values of $\omega$, they are tightly bound. 
The exact ground state solutions for the Hooke's atom 
are known for special values of $\omega$ \cite{Taut,JS}. We consider 
here the first nine values of $\omega$ for which an analytical solution 
is available: from $\omega=0.25$ (strongly bound electrons), 
to $\omega=3.597\times 10^{-6}$ (strongly-correlated electrons).
%
\begin{figure}[bt]
\includegraphics[width=\columnwidth]{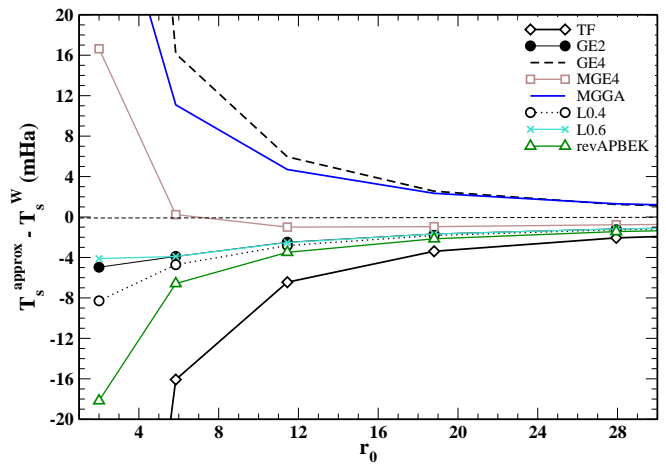}
\caption{Kinetic energy deviations from the exact values
of several KE functionals, versus the classical electron distance in 
the Hooke atom. The von Weizs\"{a}cker KE functional is the exact reference 
for a two-electron closed-shell system.} 
\label{f3}
\end{figure}
%

The MARE
with respect to the exact von Weizs\"{a}cker values 
($(T_s^{approx}-T_s^{W})/T_s^{W}\times 100$), are reported 
in Table \ref{tab1}.
The best Laplacian-level meta-GGA results are obtained with MGE4, MGGA and L0.6, which are also 
superior to the best GGAs.

In Fig. \ref{f3} we report the individual deviations for the smallest
values of the classical electron distance \cite{Taut} $r_0=(\omega^2/2)^{-1/3}$.
The L0.4 and L0.6 functionals provide the best description over the whole range
of frequencies, whereas the good performance of MGE4 originates 
from a high accuracy in the strongly correlated regime,
while in the tightly bound regime much larger errors are found.

\subsection{Jellium clusters}
\label{sec422}



%
\begin{figure}
\includegraphics[width=\columnwidth]{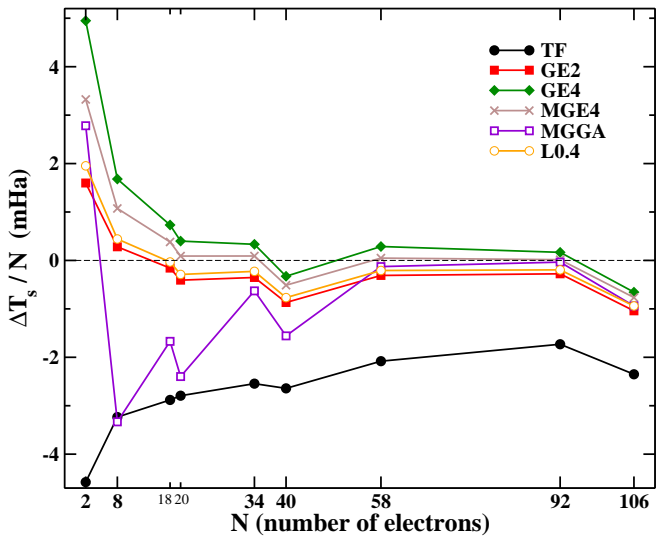}
\caption{Error on the kinetic energy per electron  
of the jellium clusters with different electron numbers. 
}
\label{f4}
\end{figure}
%
The KE MAREs for jellium clusters are reported in Table \ref{tab1}.
The best Laplacian-level meta-GGA functional is L0.4, the worst is MGGA.

The errors on the kinetic energy per electron ($\Delta T_s/N=(T_s^{approx}(N)-T_s^{exact}(N))/N$)
for the different clusters are reported
in Fig. \ref{f4} for some selected KE functionals (TF, GE2, GE4, MGE4, MGGA, L0.4).
The data in Fig. \ref{f4} show that almost of all functionals (but MGGA and TF) work very well for
medium and large clusters (i.e. for N$\ge$18 the is error below 1 mHa) whereas for smaller clusters
definitely larger errors are found.
On the other hand, the TF functional  yields always an underestimation of
the kinetic energy, less dependent on the cluster's size, whereas  
the MGGA functional shows strong oscillations.
In Fig. \ref{f4} it is worth to note the very close performances of GE2 and L0.4 for all $N$, as these two functionals
are quite different from each other (see Fig. \ref{f1}).
This small error originates from an error cancellation between the region inside the cluster
($s<1$ and  $|q|$ $<$1)
 where  L0.4 
is (by construction) almost the same of GE4 and thus larger than GE2, and the region outside the cluster 
($s$ and $q$ are large) where  GE2 is larger than L0.4, as the former diverges with $s$ while the 
latter approaches a constant (see Eq. (\ref{esasy})).

%

The dependence of the KE errors on the clusters' size suggests
that interesting results may be obtained for
the cluster disintegration problem.
We consider the disintegration of the  
cluster with $N=106$ into smaller magic clusters. The
energy associated with this process is called disintegration kinetic
energy (DKE) and is defined as \cite{PhysRevB.79.115117}
$
\text{DKE} = \sum_im_iT_s(N_i) - T_s(106) \ ,
$
with $m_i$ positive integers such that $\sum_im_iN_i=106$
(273 such processes are considered here).
If we define $\bar{T}_s(N)=T_s(N)/N$, it is easy to show that
$\text{DKE} = \sum_im_iN_i\left[\bar{T}_s(N_i)-\bar{T}_s(106)\right]\,  ,$
which shows that the errors on the DKE are obtained
as weighted sums of the differences $\bar{T}_s(N_i)-\bar{T}_s(106)$.

The DKE MARE 
are reported in Table \ref{tab1}.
It turns out that the best Laplacian-level meta-GGA functional is MGGA, which was the worst for 
the KE: this traces back to the
strong oscillations which provide a significant error cancellation.
The best overall functional is the TF functional (MARE 17.7), thanks to its almost 
uniform error among different
cluster sizes (despite it is the worst for KE, with MARE 4.4).

\subsection{Jellium surfaces and slabs}
\label{sec43}
The LDM MAREs \cite{apbekint} of several jellium surfaces
are reported in Table \ref{tab1}.
The best results are found for GE4, and in general all Laplacian-level meta-GGA functionals (recovering GE4)
overcomes the best GGAs.
In fact the uniform-electron-gas scaling becomes important in 
these systems. 

We also calculated the KE energy of two interacting jellium slabs, for various distances $0\leq z\leq \lambda_F$ 
between the slabs. The MARE represents $(1/\lambda_F)\int^{\lambda_F}_0 dz \; |T_s(z)-T_s^{exact}(z)|/T_s^{exact}(z)$, 
and is reported in Table \ref{tab1} for each functional. 
Table \ref{tab1} shows that  GE4 and all the Laplacian-level meta-GGA functionals are significantly better 
than GGAs.

Finally we also consider the following KE difference
$\text{DKE}=(1/\lambda_F)\int^{\lambda_F}_0 dz \; 
[|T_s(z)-T_s(0)|-|T_s^{exact}(z)-T_s^{exact}(0)|]/|T_s^{exact}(z)-T_s^{exact}(0)|$, 
which represents the dissociation KE of a jellium slab into two pieces 
and can be used as an indicator of the
 quality of the functionals in describing bonding regions.
Also in this case Table \ref{tab1} shows that  Laplacian-level meta-GGA functionals (GE4, L0.4 and L0.6) are significantly 
better than GGAs.

\subsection{Atoms, ions and molecules}
\label{sec44}
The KE MAREs of a benchmark set of atoms and
ions are reported in Table \ref{tab1}. 
The GGAs yield a MARE of
about 1\%, whereas all the Laplacian-level meta-GGAs are about twice worse, with a MARE in the range 1.4-2.5\%.

As an additional test, we considered the ionization kinetic energies
(IKE=$T_s^{atom}-T_s^{ion}$) of the noble gases (until Uuo). 
Note that because of the virial theorem, IKEs are just equal to the 
regular ionization potentials.
The MARE IKEs are reported in Table \ref{tab1}.
All the Laplacian-level meta-GGA functionals (but MGGA) are better than the 
GGAs.  This result traces back to the fact that the GE4-based functionals 
behave better than GGAs for the homogeneous and fractional scalings, and thus 
for systems with fluctuating number of electrons. 

Then we considered the 
total and atomization KEs (AKE) of a set of molecules.
We recall that the latter is a hard test for any kinetic energy functional
and that most of the functionals even fail to yield a qualitative description 
of AKEs \cite{PhysRevB.75.155109}.
Laplacian-level meta-GGA functionals show a MARE for the total KEs  below or close to 1\% \cite{TWESOm}, which 
is close to GE2 one but much worse than
the best GGA. 
For the AKE the trend is similar to the disintegration of the $106e^-$ jellium cluster. 

\subsection{Summary and overall assessment}
Table \ref{tab1} reports the global MARE (relative to GE2) 
for the total KE and the KE differences.
Concerning the total KE, the best Laplacian-level meta-GGA functional is L0.4, followed by L0.6.
Both functionals are largely better than GE4 (MARE reduced to one half) 
but are a little worse than the GGA functionals.
However, when KE differences are considered, the best performance is yield by the
MGGA functional, which definitely overcomes  the best GGA.
The other Laplacian-level meta-GGA functionals also show very good performances for KE difference, with 
MARE(GE2) in the range 0.93-0.96, close to the best GGA.

These results show that in general the inclusion of the Laplacian can
improve the description of the noninteracting kinetic energy. 
However, the proper dependence on this parameter is not captured in 
a systematic way by any of the functionals that we examined here.
As a result the Laplacian-level meta-GGA functionals perform in a rather erratic way
overcoming the GGAs for some properties and systems, but also
showing sudden failures for other cases.
Anyway, the results summarized in Table \ref{tab1} indicate that
the L0.4 (and L0.6) functional have a more 
regular behavior than other Laplacian-level meta-GGAs and can be competitive
in numerous applications.

\section{FDE calculations}
\label{sec5}
In this section we present the results of FDE calculations
on different test systems using the Laplacian-level meta-GGA functionals
considered in this paper, but GE4 
which gives very poor results, even failing 
to converge in some cases, possibly due to its wrong tail behavior.

\subsection{Embedding densities}
The errors on embedding densities (see Eq. (\ref{xi})) are reported in Table \ref{tab3}.
%
\begin{table*}
\begin{center}
\caption{\label{tab3} Global absolute errors on embedding densities (see Eq. (\ref{xi})), 
resulting from FDE calculations with different KE functionals on several classes of 
systems (weak, dipole, and hydrogen-bonded systems). The mean absolute error (MAE) for 
each set of molecules, and the total MAE 
are also reported. The best (worst) Laplacian-level meta-GGA value on each line is highlighted in bold 
(underline) style.
The star symbol stands for a Laplacian-level meta-GGA KE functional that is better than the best GGA.}
\begin{tabular}{lrrrrrrrrr}
\hline\hline
 &  & \multicolumn{3}{c}{GGAs} & & \multicolumn{4}{c}{meta-GGAs} \\
\cline{3-6}\cline{7-10}
 \multicolumn{1}{c}{system} &
  \multicolumn{1}{c}{TF} &
  \multicolumn{1}{c}{GE2} &
  \multicolumn{1}{c}{APBEK} &
  \multicolumn{1}{c}{revAPBEK} & &
  \multicolumn{1}{c}{MGE4} &
  \multicolumn{1}{c}{MGGA} &
  \multicolumn{1}{c}{L0.4} &
  \multicolumn{1}{c}{L0.6} \\
\hline
\multicolumn{10}{c}{\bf{Weak interaction}} \\
He-Ne   &    0.71 & 0.93 & 0.71 & 0.70 & & 1.46 & \underline{1.53} & \textbf{0.76} & \textbf{0.76} \\
He-Ar   &    0.78 & 1.16 & 0.78 & 0.78 & & 1.88 & \underline{2.01} & 0.85 & \textbf{0.83} \\
Ne-Ne   &    0.17 & 0.50 & 0.12 & 0.08 & & \underline{1.58} & \underline{1.58} & 0.26 & \textbf{0.21} \\
Ne-Ar   &    0.22 & 0.65 & 0.14 & 0.10 & & \underline{2.02} & \underline{2.02} & 0.31 & \textbf{0.19} \\
CH$_4$-Ne &  0.28 & 0.75 & 0.17 & 0.12 & & \underline{2.54} & 2.53 & 0.39 & \textbf{0.29} \\
C$_6$H$_6$-Ne&0.41& 0.64 & 0.19 & 0.18 & & \underline{2.32} & \underline{2.32} & 0.38 & \textbf{0.21} \\
CH$_4$-CH$_4$&0.59 &1.39 & 0.29 & 0.60 & & \underline{5.89} & 5.64 & \textbf{0.36} & 0.77 \\
        &         &      &      &      & &      &      &      &   \\
MAE     &    0.45 & 0.86 & 0.34 & 0.37 & & \underline{2.53} & 2.52 & \textbf{0.47} & \textbf{0.47} \\
\hline
\multicolumn{10}{c}{\bf{Dipole-dipole interaction}} \\
H$_2$S-H$_2$S & 2.19 & 2.29 & 2.01 & 2.08 & & \textbf{2.17} & \underline{4.67} & 2.73 & 2.86 \\
HCl-HCl       & 2.70 & 2.60 & 2.47 & 2.50 & & \textbf{2.54} & \underline{4.56} & 2.76 & 2.86 \\
H$_2$S-HCl    & 5.07 & 4.64 & 4.79 & 4.78 & & \textbf{4.17} & \underline{6.59} & 4.73 & 4.77 \\
CH$_3$Cl-HCl  & 3.35 & 3.25 & 3.04 & 3.08 & & 4.07 & \underline{5.35} & \textbf{3.15} & 3.28 \\
CH$_3$SH-HCN  & 1.86 & 2.29 & 1.91 & 2.05 & & 3.26 & \underline{5.08} & \textbf{2.56} & 2.68 \\
CH$_3$SH-HCl  & 5.47 & 5.22 & 5.22 & 5.24 & & 5.71 & \underline{6.60} & \textbf{5.13} & 5.15 \\
              &      &      &      &      & &      &      &      &   \\
MAE           & 3.44 & 3.38 & 3.24 & 3.29 & & 3.65 & \underline{5.47} & \textbf{3.51} & 3.60 \\
\hline  
\multicolumn{10}{c}{\bf{Hydrogen-bond interaction}} \\
NH$_3$-NH$_3$  & 2.32 & 2.30 & 2.04 & 2.12 & & \textbf{2.50} & \underline{5.44} & 2.92 & 2.91 \\
HCl-HCl        & 2.69 & 2.38 & 2.22 & 2.19 & & \textbf{2.28} & \underline{5.36} & 2.35 & 2.36 \\
H$_2$O-H$_2$O  & 2.94 & 2.73 & 2.55 & 2.58 & & \textbf{2.67} & \underline{6.02} & 3.02 & 2.96 \\
NH$_3$-H$_2$O  & 4.57 & 4.14 & 4.26 & 4.26 & & \textbf{4.04} & \underline{7.38} & 4.52 & 4.47 \\
HF-NCH         & 4.49 & 4.22 & 4.22 & 4.22 & & \textbf{4.09} & \underline{6.80} & 4.29 & 4.25 \\
(HCONH$_2$)$_2$& 3.17 & 3.32 & 3.17 & 3.30 & & *\textbf{3.13} & \underline{6.78} & 3.92 & 3.78 \\
(HCOOH)$_2$    & 4.84 & 4.54 & 4.65 & 4.69 & & *\textbf{4.03} & \underline{6.85} & 4.82 & 4.69 \\
               &      &      &      &      & &      &      &      &   \\
MAE            & 3.57 & 3.38 & 3.30 & 3.34 & & *\textbf{3.25} & \underline{6.38} & 3.69 & 3.63 \\
\hline
\multicolumn{10}{c}{\bf{Overall assessment}} \\
MAE            & 1.81 & 1.85 & 1.66 & 1.69 & & 2.31 & \underline{3.52} & \textbf{1.86} & \textbf{1.86} \\
\hline\hline
\end{tabular}
\end{center}
\end{table*}
%
We recall that this is an important test for embedding approaches
because it provides direct insight into the quality of the 
embedding potential 
\cite{laricchia10,apbekint,Laricchia2011114,beyhan10,jacob08prot,neugebauer08,Govind09cpl,fux10}. 

Inspection of the data shows that the L0.4 and L0.6
functionals perform very similarly and are in line with
the GGA functionals. 
On the other hand, MGE4 and MGGA provide significantly larger errors.
For weakly interacting systems both functionals display 
a very poor performance and interestingly almost the same results.
This finding is rationalized considering that, for 
dispersion-dominated, systems the bond region
is characterized by small and medium values of $s$ and quite large
values of $q$ ($q\gtrsim5$). Thus, in this region
the two functionals, and the corresponding kinetic potentials, 
are the same by construction (in this region $F_s^{MGE4}>F_S^{W}$, hence
$F_s^{MGGA}=F_s^{MGE4}$; see Eqs. (\ref{pr6}) and (\ref{pr4}).
On the other hand, for the other kinds of interactions
only MGGA shows a poor behavior whereas MGE4 performs rather well, 
being even the best Laplacian-level meta-GGA for hydrogen bonds. 
In this cases in fact the bonding region is characterized by
 small and medium values of $s$ but small values of 
$q$ ($q\sim 0.5$) so that MGE4 recovers GE4 which is a reasonable 
limit. However, this is exactly the range of values 
where the sharp interpolating function of MGGA assumes
its intermediate values. Therefore, the MGGA potential
is strongly oscillating in this region (see Fig. \ref{f1}).

\begin{figure}
\includegraphics[width=\columnwidth]{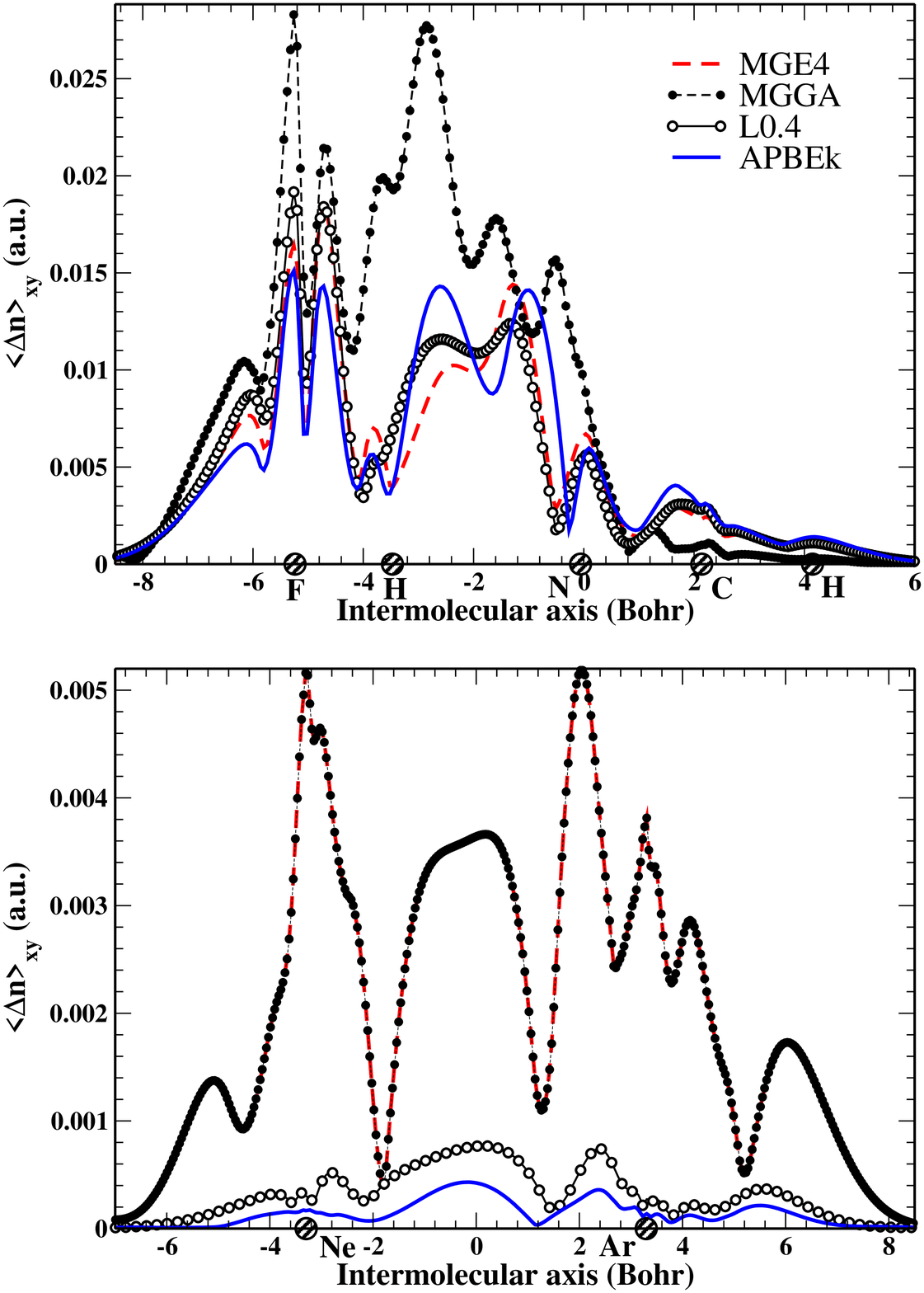}
\caption{Plane averaged deformation densities obtained from 
different KE functionals, for the HF-NCH hydrogen-bond complex (upper panel) 
and the weakly-interacting Ne-Ar dimer. 
The filled circles on the $x$ axis denote the atoms' positions.} 
\label{f7}
\end{figure}

To understand better these results we consider in Fig. (\ref{f7})
the plot of the plane-averaged deformation density in two
typical cases: for the hydrogen-bond complex HF-NCH and for the
weakly-interacting Ne-Ar system.
%
%
The figure confirms the findings of Table \ref{tab3} and
additionally indicates that the GGA and L0.4/L0.6 densities
are in fact similar over the whole space (not only after integration).
On the contrary, MGE4 is very similar to L0.4 for hydrogen-bond
complexes, but almost identical to MGGA for weakly interacting complexes.
Note that the HF-NCH complex has
larger density in the bond than Ne-Ar, thus the relative errors in the
 bond are larger in the latter case that in the former.

\subsection{Embedding energies}
\label{sec:enerr}
The errors on the embedding energies obtained from FDE calculations
using different kinetic functionals are reported in Table \ref{tab4}.
%
\begin{table*}
\begin{center}
\caption{\label{tab4} Embedding energy errors (see Eq. (\ref{deltaee})) 
in mHa, resulting from FDE calculations with different KE functionals on 
several classes of systems (weak, dipole, and hydrogen-bonded systems). 
The second column reports the benchmark binding energy $E_b$ from Ref. \onlinecite{truhlar05nb}. 
The mean error (ME), MAE, and MARE are indicated for each set of molecules. 
At the bottom of the table, the total MAEs are  
also reported. 
The best (worst) Laplacian-level meta-GGA value on each line is highlighted in bold (underline) style.
The star symbol stands for a Laplacian-level meta-GGA KE functional that is better than the best GGA.}
\begin{small}
\begin{tabular}{lrrrrrrrrrr}
\hline\hline
 &  & &  \multicolumn{3}{c}{GGAs} & & \multicolumn{4}{c}{meta-GGAs} \\
\cline{4-7}\cline{8-11}
 \multicolumn{1}{c}{system} &
 \multicolumn{1}{c}{$E_b$} &
  \multicolumn{1}{c}{TF} &
  \multicolumn{1}{c}{GE2} &
  \multicolumn{1}{c}{APBEK} &
  \multicolumn{1}{c}{revAPBEK} & &
  \multicolumn{1}{c}{MGE4} &
  \multicolumn{1}{c}{MGGA} &
  \multicolumn{1}{c}{L0.4} &
  \multicolumn{1}{c}{L0.6} \\
\hline
\multicolumn{11}{c}{\bf{Weak interaction}} \\
He-Ne   &    0.06 & 0.38 & -0.81 & 0.39 & 0.35 & & \underline{-3.28} & -3.24 & \textbf{0.48} & 0.50 \\
He-Ar   &    0.10 & 0.43 & -0.81 & 0.44 & 0.40 & & \underline{-3.20} & -3.13 & \textbf{0.55} & 0.57 \\
Ne-Ne   &    0.13 & 0.29 & -1.67 & 0.27 & 0.18 & & \underline{-4.45} & \underline{-4.45} & \textbf{0.50} & 0.52 \\
Ne-Ar   &    0.21 & 0.34 & -1.84 & 0.29 & 0.16 & & \underline{-4.44} & \underline{-4.44} & \textbf{0.57} & \textbf{0.57} \\
CH$_4$-Ne &  0.35 & 0.37 & -2.05 & 0.32 & 0.18 & & \underline{-7.11} &\underline{ -7.11} & \textbf{0.63} & 0.64 \\
C$_6$H$_6$-Ne& 0.75&1.05 & -4.49 & 0.61 & 0.14 & & \underline{-13.39} & \underline{-13.39} & 1.59 & \textbf{1.45} \\
CH$_4$-CH$_4$& 0.81 & 0.78 & -4.39 & 0.21 & -0.29 & & \underline{-13.29} & \underline{-13.13} & 1.05 & \textbf{0.67} \\
        &    &     &      &      &      & &      &      &      &   \\
MAE     &         & 0.52 & 2.29 & 0.36 & 0.24 & & \underline{7.02} & 6.98 & 0.77 & \textbf{0.70} \\
\hline
\multicolumn{11}{c}{\bf{Dipole-dipole interaction}} \\
H$_2$S-H$_2$S &  2.63 & 2.15 & -4.68 & 0.67 & -0.23 & & -3.83 & \underline{-7.80} & *\textbf{-0.04} & -1.19 \\
HCl-HCl       &  3.20 & 3.26 & -4.76 & 1.49 & 0.44 & & -3.67 & \underline{-8.76} & 0.73 & \textbf{-0.60} \\
H$_2$S-HCl    &  5.34 & 5.28 & -4.83 & 2.46 & 1.05 & & -3.91 & \underline{-12.55} & *\textbf{0.44} & *-1.03 \\
CH$_3$Cl-HCl  &  5.66 & 5.92 & -7.08 & 2.42 & 0.63 & & -7.37 & \underline{-15.40} & 1.12 & \textbf{-0.82} \\
CH$_3$SH-HCN  &  5.72 & 2.79 & -7.34 & 0.53 & -0.79 & & -9.84 & \underline{-16.55} & *\textbf{-0.24} & -1.78 \\
CH$_3$SH-HCl  &  6.63 & 8.46 & -6.30 & 3.80 & 1.72 & & -7.23 & \underline{-19.07} & *1.24 & *\textbf{-0.65} \\
              & &     &      &      &      & &      &      &      &   \\
MAE           &       & 4.64 & 5.83 & 1.90 & 0.81 & & 5.97 & \underline{13.36} & *\textbf{0.64} & 1.01 \\
\hline  
\multicolumn{11}{c}{\bf{Hydrogen-bond interaction}} \\
NH$_3$-NH$_3$  &  5.02 & 2.63 & -6.20 & 0.63 & -0.55 & & -5.21 & \underline{-11.44} & *\textbf{-0.50} & -1.94 \\
HCl-HCl        &  7.28 & 5.23 & -5.87 & 2.77 & 1.29 & & -4.21 & \underline{-15.32} & 1.42 & \textbf{-0.49} \\
H$_2$O-H$_2$O  &  7.92 & 4.77 & -6.82 & 1.84 & 0.24 & & -5.42 & \underline{-16.22} & *\textbf{-0.17} & -1.93 \\
NH$_3$-H$_2$O  &  10.21 & 5.97 & -6.98 & 2.17 & 0.34 & & -6.28 & \underline{-19.97} & \textbf{-0.86} & -2.65 \\
HF-NCH         &  11.33 & 8.26 & -6.62 & 3.78 & 1.67 & & -5.85 & \underline{-23.46} & *\textbf{0.17} & -1.89 \\
(HCONH$_2$)$_2$&  23.81 & 10.59 & -17.72 & 1.43 & -2.76 & & -17.85 & \underline{-52.06} & \textbf{-5.42} & -9.11 \\
(HCOOH)$_2$    &  25.74 & 18.81 & -15.77 & 6.13 & 0.99 & & -15.49 & \underline{-60.77} & \textbf{-3.49} & -7.26 \\
               &   &   &      &      &      & &      &      &      &   \\
MAE            &       & 8.04 & 9.43 & 2.68 & 1.12 & & 8.62 & \underline{28.46} & \textbf{1.72} & 3.61 \\
\hline
\multicolumn{11}{c}{\bf{Overall assessment}} \\
MAE            &     & 4.39 & 5.85 & 1.63 & 0.72 & & 7.27 & \underline{16.41} & \textbf{1.06} & 1.81 \\
\hline\hline
\end{tabular}
\end{small}
\end{center}
\end{table*}
%
The data show that the L0.4 functional has $\text{MAE}= 1.06$ 
mHa, which is 
comparable to that of the state-of-the-art GGA KE functionals and lower 
than the errors originating from the XC approximation \cite{truhlar05nb}.
In particular, L0.4 yields the best results ($\text{MAE}=0.64$ mHa) 
for the dipole-dipole interaction systems, 
also outperforming revAPBEK.
On the other hand, lower accuracy is obtained for the weakly 
interacting systems: this drawback can be 
related to the 
inaccuracy of the gradient expansion for this class of systems, as it will be explained below (see next section).
In general, we can state the important result that the L0.4 Laplacian-level
functional can be effectively used to approximate the non-additive kinetic energy 
functional in embedding calculations, yielding accurate total embedding energies 
for non-covalently interacting systems.

Concerning the other Laplacian-level meta-GGA functionals, slightly worse results are found with L0.6,
which gives in any case rather good results, in line with APBEK.
For the MGE4 and MGGA functionals, similar considerations as 
for the case of the embedding densities apply.
In fact, as shown in more details in next section, for weakly-interacting systems the embedding energy 
is mainly determined by the region having moderately large $s$-values
and large values of the reduced Laplacian ($q\gtrsim 5$).
Hence, the two functionals perform very similarly and yield strongly
underestimated embedding energies (see next section).
For dipole-dipole and hydrogen bond interactions, instead, 
the relevant region for the embedding energy is defined (see next section) 
by  relatively small values of the reduced parameters ($s\lesssim 1$ and
$|q|\lesssim 1$). Thus, MGE4 correctly tends to GE4, which is a rather
good approximation for this limit, whereas MGGA is dominated by
the interpolating function. 

\subsection{Energy decomposition analysis}

To analyze in more details 
Laplacian-level kinetic energy functionals, we extend the 
idea proposed in Ref. \onlinecite{apbekint}.
Thus, we perform a decomposition of the non-additive
KE in terms of the reduced gradient and Laplacian
contributions.
Similarly with the GGA case \cite{apbekint}, we define the following transformation
of the Thomas-Fermi kinetic energy density
\begin{equation}
t[n](s,q)=\int \tau^{TF}[n](\R)\delta(s-s(\R))\delta(q-q(\R)) \, d\R \ ,
\label{eq:deft}
\end{equation}
so that the Thomas-Fermi kinetic energy is
\begin{equation}
T_s^{TF}[n]=\int\int t[n](s,q) \, ds\, dq,
\end{equation}
where $t[\rho](s,q)$ is the $(s,q)$-decomposed homogeneous electron gas (HEG)
KE distribution. 
For a Laplacian-level meta-GGA KE functional we have
\begin{equation}
T_s[F_s,n]=\int\int t[n](s,q) F_s(s,q) \, ds\, dq \label{eq:sts}\, .
\end{equation}
Equation (\ref{eq:sts}) states that the total kinetic energy 
is the scalar product (in the $(s,q)$-space) of $t(s,q)$ and the kinetic 
enhancement factor $F_s(s,q)$. Thus,
in this formalism $T_s$ is also a functional of $F_s$.
Then, as the definition of the non-additive kinetic energy
is linear in the composing total KEs (see Eq. (\ref{eq:kinn})), we
also obtain
\begin{equation}\label{eq:sd}
T_s^{nadd}[F_s;n_\A;n_\B]=\int \int t^{nadd}[n_\A;n_\B](s,q) F_s(s,q) \, ds\, dq\ ,
\end{equation}
with
\begin{align}
t^{nadd}[n_\A;n_\B](s,q)&=t[n_\A+n_\B](s,q) +\nonumber \\
                           &-t[n_\A](s,q)-t[n_\B](s,q) \ .
\label{eq101}
\end{align}
%
Equation \ref{eq101}
is a generalization of Eq. (19) of Ref. \onlinecite{apbekint}.
Note that 
the ($s,q$)-decomposition is a more powerful tool as compared to the  $s$-decomposition, because the former 
can distinguish  important density regions:  
$s\approx 0$, $q>0$ in the bond, 
$s\leq 0.4$, $q<0$ near the nucleus, which cannot be resolved in the  $s$-only decomposition.

Following Ref. \onlinecite{apbekint}, we thus have that the embedding energy error, 
for given approximated embedded density, can be written as:
\begin{equation}\label{qwe}
\Delta E[F_s] \approx \int \int t^{nadd}[n_\A;n_\B](s,q) F_s(s,q) \, ds \, dq + \Delta W
\end{equation}
where  $\Delta W$ is a constant, i.e. a known bifunctional of the embedded densities.

We note that, if a GGA enhancement factor is considered, 
using $\int_{-\infty}^{\infty} dx\; \delta(x)=1$,
Eq. (\ref{qwe}) correctly turns into the GGA expression
\begin{eqnarray}
\nonumber
\Delta E[F_s] & \approx & \int F_s(s) \left[\int t^{nadd}[{n}_\A;{n}_\B](s,q)  
dq\right]ds   + \Delta W \\
& = &\int F_s(s) \langle{t}^{nadd}[{n}_\A;{n}_\B]\rangle_q(s)ds\ + \Delta W ,
\end{eqnarray}
i.e. the $q$-averaged integral non-additive $s,q$-decomposed HEG kinetic energy distribution
$\langle{t}^{nadd}[{n}_\A;{n}_\B]\rangle_q(s)$ corresponds to the
non-additive HEG kinetic energy distribution obtained in Ref. \onlinecite{apbekint}
for the $s$-decomposition of a GGA functional 
(see Eqs. (14) and (19) of Ref. \onlinecite{apbekint}).

Equation (\ref{qwe}) can be used to understand the role of $F_s(s,q)$
and the performance of different
Laplacian-level meta-GGAs in terms of the shape of their
enhancement factor in the $(s,q)$-space.

As an application of the present  ($s,q$)-decomposition we consider the 
HF-NCH and Ne$_2$ complexes. The L0.4 functional gives almost the exact energy for 
the former (embedding error of only 0.17mHa)
whereas it is quite inaccurate for the latter (embedding error 
larger than the binding energy).
%
\begin{figure*}
\includegraphics[width=\textwidth]{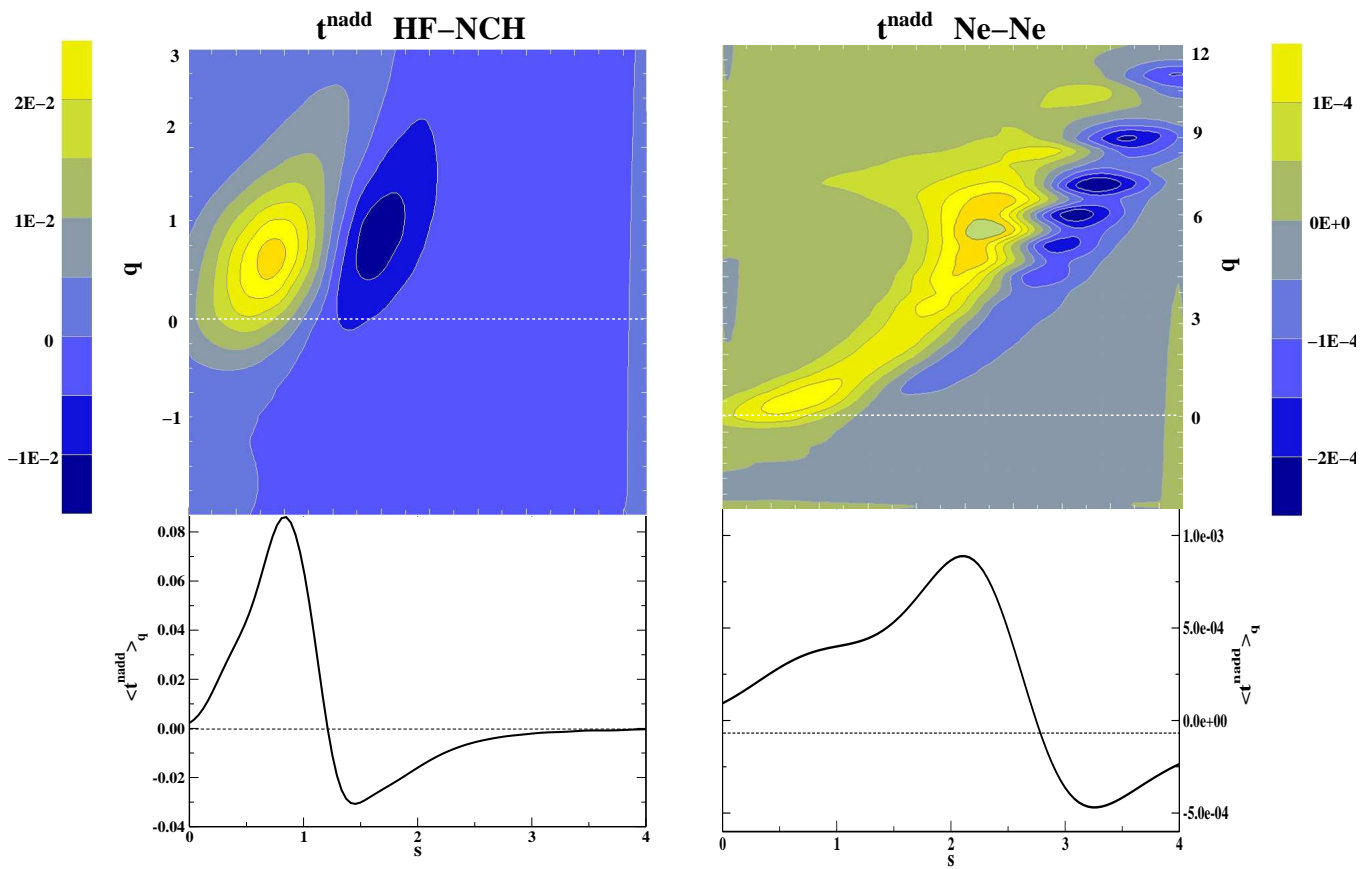}
\caption{
$s,q$-decomposed non-additive HEG kinetic energy 
distribution ($t^{nadd}(s,q)$) for the complex HF-NCH (left) and for Ne$_2$ (right), with $s$ and $q$ in the
range appropriate to physical densities. The corresponding $q$-averaged integral
$t^{nadd}(s,q)$ 
($\braket{t^{nadd}(s)}_q$) is also reported for comparison in the bottom panels. 
} 
\label{f8}
\end{figure*}
%
Figure \ref{f8} reports, for the two systems, the non-additive $(s,q)$-decomposed HEG
kinetic energy distribution $t^{nadd}(s,q)$  (upper-panel),
as well as the plot of
$\langle t^{nadd}\rangle_q(s)$
(lower panel), both calculated by fixing the embedded densities to the APBEK ones.
Integration in Eq. (\ref{eq:deft}) was  performed numerically, representing
the delta function with a Gaussian model with broadening
$\sigma$=0.07.

The plots on the bottom panels show that, in analogy with the
GGA case studied in Ref. \onlinecite{apbekint},
the correct embedding energy comes from
a delicate balancing of positive contributions at
small $s$ values and negative contributions at
larger values of $s$ in the scalar product (Eq. (\ref{eq:sd})). 
On the other hand, richer structures are present considering $t^{nadd}(s,q)$ and several considerations 
can be done:
\par
i) Contributions from negative $q$ are negligible, 
meaning that the core regions are not important for  
$t^{nadd}(s,q)$.
This information cannot be extracted from the $s$-only decomposition because both 
the core region as well as the valence region have  $s\lessapprox0.4$. 
\par
ii) for HF-NCH the $q$-dependence of $t^{nadd}(s,q)$ is quite weak, 
explaining the success of GGA approximations,
which relay on the $q$-averaged integral of $t^{nadd}(s,q)$. Moreover,
the non-zero values of  $t^{nadd}(s,q)$ are mostly confined in the range $0 \le q \le 2$ and $0 \le s 
\le 2$, which is related to the bonding region
(the value of $q$ at the center of the bond is $q\approx 0.5$). 
In this region of the $s,q$-space (small $s$ and small $q$) 
the fourth-order gradient expansion can be correct, explaining the very low embedding errors
of L0.4 (which recovers GE4).
Similar plots have been obtained for other hydrogen-bond or dipole-dipole 
interaction systems.
\par
iii) For Ne$_2$, the $t^{nadd}(s,q)$ looks very different. The most important 
structures are now at $s > 1.5$ and
 $4 \le q \le 8$. In fact, due to the weakly interacting character of the system, 
the value of the reduced Laplacian at the center
of the bond is $q\approx 6.8$. The plot thus shows how difficult can be the construction of an accurate 
Laplacian-level kinetic functional for FDE.
In fact, the embedding energy error depends on the product over the whole $(s,q)$-space of  
$t^{nadd}(s,q)$ and the kinetic enhancement factor $F_s(s,q)$.
Hence, an accurate enhancement factor should properly take into account the 
complexity of the structures at $4 \le q \le 8$ and $s > 1.5$. However, the L0.4/L0.6 
enhancement factor is only weakly dependent on $q$ for $s>3$.
On the other hand, GGA functionals average over $q$ and loose all $q$-dependent structures: 
nevertheless, they can still be very accurate for weakly-interaction systems (see e.g. revAPBEK) thanks 
to an error cancellation in different $q$-regions.
This error cancellation is however less likely (and also undesired) for Laplacian-level kinetic 
functionals. 

Finally, the plot of Fig. \ref{f8} also explains the fact
that MGGA and MGE4 always yield strongly underestimated
embedding energies. This fact traces back to the
diverging enhancement factor as $s$ increases
(see Fig. \ref{f1}).
This {\it exact} feature brings an overestimation of the
negative contributions of $t^{nadd}(s,q)$ (always located at large $s$ values)
which is not well balanced by the regions responsible for the 
positive contributions. 
Thus, too negative embedding energies 
(even more negative than for GE2) are obtained.

\section{Conclusions}
\label{sec:conclusion}
In this work we investigated the significance of the fourth-order gradient
expansion of the kinetic energy. To this end we performed a throughout
assessment of several Laplacian-level meta-GGA kinetic energy functionals,
with special attention to subsystem DFT applications.

Our study indicated that the inclusion of informations coming from the Laplacian of the
density into the functional may play an important role towards a 
higher accuracy and broader applicability.
In fact, GE4 significantly improves over GE2 for 
solid-state related models (as jellium
surfaces, interacting jellium slabs, and large jellium clusters;
 see Table \ref{tab1}) as well as for the monovacancy formation
in jellium \cite{monov}.
Nevertheless, GE4 shows serious drawbacks for small finite systems (e.g. light atoms), due to
a non-physical behavior near the nucleus and in the tail of the density, which make the
construction of GE4-based functionals a real challenge.
In fact, the different Laplacian-level meta-GGAs tested in this paper displayed a
quite disappointing unsystematic accuracy, being quite good
for some properties and systems and rather poor (at least worse than
GGA methods) for others.
These problems become especially evident in FDE applications, where
the quality of the nonadditive kinetic potential plays a fundamental role,
being applied to different densities at the same time.
In fact, most of the Laplacian-level meta-GGA functionals considered in the
present study perform poorly and several ones also yield severe
convergence problems.

The notable exception to this
behavior is given by the L0.4 and L0.6 functionals,
which perform overall rather close to the best GGAs,
especially in embedding calculations of small molecular complexes, while 
being better than the best GGAs for solid-state related jellium models. 
The reason for this traces back to the fact that these functionals
were constructed not only to recover GE4 in the slowly-varying
density limit, but also to achieve a reasonably good behavior
in the rapidly-varying regime. This later goal was obtained
by mimicking the successful behavior of the (rev)APBEK functional
in the rapidly-varying limit.
Thus, the L0.4 and L0.6 functionals,  
appear as promising tools
for the application of Laplacian-level meta-GGA kinetic energy functionals
in the context of FDE or the density-decomposed orbital-free DFT 
\cite{CarterXia,PhysRevB.85.045126}. 
 
We remark, however, that the main message emerging from the present work is
that there is still a huge amount of work to be done
in the development of Laplacian-level meta-GGA kinetic energy functionals 
before they can significantly overcome the more simple GGAs 
for FDE applications to weakly interacting systems.
This future work should be focused on studying in deeper details the
role played by the Laplacian in different systems and density regimes, so that
more complex dependences on the $q$ variable can be developed,
especially in the rapidly-varying density regime.
In fact, with our assessment work, and in particular through the
$(s,q)$-decomposition technique, we showed that the actual Laplacian-level meta-GGA functionals
display a reliable $q$-dependence only in the slowly-varying limit,
whereas they show limitations in the rapidly-varying regions.

\begin{acknowledgment}
This work was partially funded by the European Research Council (ERC) 
Starting Grant FP7 Project DEDOM, Grant No. 207441. The authors thank 
TURBOMOLE GmbH for providing the TURBOMOLE program package and M. 
Margarito for technical support.
\end{acknowledgment}

\appendix
\section{Implementation of Laplacian-level meta-GGAs}
For any Laplacian level DFT functional of the form
\begin{equation}
T_s[n] = \int \tau_s[n,\nabla n,\nabla^2n](\R)d\mathbf{r}
\end{equation}
the functional derivative with respect to the electron density is:
\begin{equation}
\frac{\delta T_s[n]}{\delta n(\mathbf{r})} =
 \frac{\partial \tau_s[n](\R)}{\partial n(\R)} 
- \nabla\cdot\frac{\partial \tau_s[n](\R)}{\partial \nabla n(\R)} 
+ \nabla^2\frac{\partial\tau_s[n](\R)}{\partial \nabla^2 n(\R)}\ .
\end{equation}
The matrix elements between the basis set functions $\{\chi_i\}$ 
required in FDE calculations are therefore
\begin{eqnarray}
\left(v_s\right)_{\mu\nu} & = &
 \int\chi_\mu(\mathbf{r})\frac{\delta T_s[n]}{\delta n(\mathbf{r})}\chi_\nu(\mathbf{r})d\mathbf{r}  \nonumber \\
&=& \int\chi_\mu(\mathbf{r})\frac{\partial \tau_s[n](\R)}{\partial n(\R)}(\mathbf{r})\chi_\nu(\mathbf{r})d\mathbf{r}   \nonumber \\ 
&+& \int \frac{\partial \tau_s[n](\R)}{\partial \nabla n}(\mathbf{r})\cdot \left[ \nabla\chi_\mu(\mathbf{r})\chi_\nu(\mathbf{r})    
 +\nabla\chi_\nu(\mathbf{r})\chi_\mu(\mathbf{r}) \right] d\mathbf{r}  \nonumber \\  
&+& \int \frac{\partial\tau_s[n](\R)}{\partial \nabla^2 n(\R)}(\mathbf{r})\big [
     \nabla^2\chi_{\mu}(\mathbf{r})\chi_\nu(\mathbf{r})
     +2\nabla\chi_{\mu}(\mathbf{r})\cdot\nabla\chi_{\nu}(\mathbf{r})  \nonumber \\  
     &+&  \chi_{\mu}(\mathbf{r})\nabla^2\chi_{\nu}(\mathbf{r})
                                                            \big]d\mathbf{r}
\end{eqnarray}
where we used the first 
($\int f(\mathbf{r})\nabla\cdot \mathbf{v}(\mathbf{r})d\mathbf{r} = -\int\nabla 
 f(\mathbf{r})\cdot \mathbf{v}(\mathbf{r})d\mathbf{r}$)
and the second Green's idenity
($\int f(\mathbf{r})\nabla^2g(\mathbf{r})d\mathbf{r}=\int\nabla^2f(\mathbf{r})
g(\mathbf{r})d\mathbf{r}$).
%


\begin{mcitethebibliography}{136}
\providecommand*\natexlab[1]{#1}
\providecommand*\mciteSetBstSublistMode[1]{}
\providecommand*\mciteSetBstMaxWidthForm[2]{}
\providecommand*\mciteBstWouldAddEndPuncttrue
  {\def\EndOfBibitem{\unskip.}}
\providecommand*\mciteBstWouldAddEndPunctfalse
  {\let\EndOfBibitem\relax}
\providecommand*\mciteSetBstMidEndSepPunct[3]{}
\providecommand*\mciteSetBstSublistLabelBeginEnd[3]{}
\providecommand*\EndOfBibitem{}
\mciteSetBstSublistMode{f}
\mciteSetBstMaxWidthForm{subitem}{(\alph{mcitesubitemcount})}
\mciteSetBstSublistLabelBeginEnd
  {\mcitemaxwidthsubitemform\space}
  {\relax}
  {\relax}

\bibitem[Thomas(1926)]{thomas26}
Thomas,~L.~H. The calculations of atomic fields. \emph{Proc. Cambridge Phil.
  Soc.} \textbf{1926}, \emph{23}, 542\relax
\mciteBstWouldAddEndPuncttrue
\mciteSetBstMidEndSepPunct{\mcitedefaultmidpunct}
{\mcitedefaultendpunct}{\mcitedefaultseppunct}\relax
\EndOfBibitem
\bibitem[Fermi(1927)]{fermi27}
Fermi,~E. Un metodo statistico per la determinazione di alcune proprieta'
  dell'atomo. \emph{Rend. Accad. Naz. Lincei} \textbf{1927}, \emph{6},
  602\relax
\mciteBstWouldAddEndPuncttrue
\mciteSetBstMidEndSepPunct{\mcitedefaultmidpunct}
{\mcitedefaultendpunct}{\mcitedefaultseppunct}\relax
\EndOfBibitem
\bibitem[Fermi(1928)]{fermi28}
Fermi,~E. A statistical method for the determination of some atomic properties
  and the application of this method to the theory of the periodic system of
  elements. \emph{Z. Phys.} \textbf{1928}, \emph{48}, 73\relax
\mciteBstWouldAddEndPuncttrue
\mciteSetBstMidEndSepPunct{\mcitedefaultmidpunct}
{\mcitedefaultendpunct}{\mcitedefaultseppunct}\relax
\EndOfBibitem
\bibitem[Parr and Yang(1989)Parr, and Yang]{dftbook}
Parr,~R.~G.; Yang,~W. \emph{Density-Functional Theory of Atoms and Molecules};
  Oxford University Press: Oxford, 1989\relax
\mciteBstWouldAddEndPuncttrue
\mciteSetBstMidEndSepPunct{\mcitedefaultmidpunct}
{\mcitedefaultendpunct}{\mcitedefaultseppunct}\relax
\EndOfBibitem
\bibitem[Dreizler and Gross(1990)Dreizler, and Gross]{dftbookgross}
Dreizler,~R.~M.; Gross,~E. K.~U. \emph{Density Functional Theory}; Springer:
  Heidelberg, 1990\relax
\mciteBstWouldAddEndPuncttrue
\mciteSetBstMidEndSepPunct{\mcitedefaultmidpunct}
{\mcitedefaultendpunct}{\mcitedefaultseppunct}\relax
\EndOfBibitem
\bibitem[Lign\`eres and Carter(2005)Lign\`eres, and Carter]{carterrev}
Lign\`eres,~V.~L.; Carter,~E.~A. \emph{Handbook of Materials Modeling};
  Springer Science and Business Media: Netherlands, 2005; pp 137--148\relax
\mciteBstWouldAddEndPuncttrue
\mciteSetBstMidEndSepPunct{\mcitedefaultmidpunct}
{\mcitedefaultendpunct}{\mcitedefaultseppunct}\relax
\EndOfBibitem
\bibitem[Chen and Zhou(2008)Chen, and Zhou]{chen08}
Chen,~H.; Zhou,~A. Orbital-free density functional theory for molecular
  structure calculations. \emph{Numer. Math. Theor. Meth. Appl.} \textbf{2008},
  \emph{1}, 1--28\relax
\mciteBstWouldAddEndPuncttrue
\mciteSetBstMidEndSepPunct{\mcitedefaultmidpunct}
{\mcitedefaultendpunct}{\mcitedefaultseppunct}\relax
\EndOfBibitem
\bibitem[{von Weizs\"{a}cker}(1935)]{vW}
{von Weizs\"{a}cker},~C.~F. Zur Theorie der Kernmassen. \emph{Z. Phys. A}
  \textbf{1935}, \emph{96}, 431--458\relax
\mciteBstWouldAddEndPuncttrue
\mciteSetBstMidEndSepPunct{\mcitedefaultmidpunct}
{\mcitedefaultendpunct}{\mcitedefaultseppunct}\relax
\EndOfBibitem
\bibitem[Kirzhnitz(1957)]{Kirz57}
Kirzhnitz,~D. Quantum corrections to the Thomas--Fermi equation. \emph{Sov.
  Phys. JETP} \textbf{1957}, \emph{5}, 64\relax
\mciteBstWouldAddEndPuncttrue
\mciteSetBstMidEndSepPunct{\mcitedefaultmidpunct}
{\mcitedefaultendpunct}{\mcitedefaultseppunct}\relax
\EndOfBibitem
\bibitem[Yonei and Tomishima(1965)Yonei, and Tomishima]{Tomishina65}
Yonei,~K.; Tomishima,~Y. On the Weizs\"{a}cker correction to the Thomas--Fermi
  theory of the atom. \emph{J. Phys. Soc. Jpn} \textbf{1965}, \emph{20},
  1051\relax
\mciteBstWouldAddEndPuncttrue
\mciteSetBstMidEndSepPunct{\mcitedefaultmidpunct}
{\mcitedefaultendpunct}{\mcitedefaultseppunct}\relax
\EndOfBibitem
\bibitem[Oliver and Perdew(1979)Oliver, and Perdew]{OP79}
Oliver,~G.~L.; Perdew,~J.~P. Spin--density gradient expansion for the kinetic
  energy. \emph{Phys. Rev. A} \textbf{1979}, \emph{20}, 397--403\relax
\mciteBstWouldAddEndPuncttrue
\mciteSetBstMidEndSepPunct{\mcitedefaultmidpunct}
{\mcitedefaultendpunct}{\mcitedefaultseppunct}\relax
\EndOfBibitem
\bibitem[Murphy(1981)]{Murphy81}
Murphy,~D.~R. Sixth--order term of the gradient expansion of the
  kinetic--energy density functional. \emph{Phys. Rev. A} \textbf{1981},
  \emph{24}, 1682--1688\relax
\mciteBstWouldAddEndPuncttrue
\mciteSetBstMidEndSepPunct{\mcitedefaultmidpunct}
{\mcitedefaultendpunct}{\mcitedefaultseppunct}\relax
\EndOfBibitem
\bibitem[Wigner(1932)]{PhysRev.40.749}
Wigner,~E. On the quantum correction for thermodynamic equilibrium. \emph{Phys.
  Rev.} \textbf{1932}, \emph{40}, 749--759\relax
\mciteBstWouldAddEndPuncttrue
\mciteSetBstMidEndSepPunct{\mcitedefaultmidpunct}
{\mcitedefaultendpunct}{\mcitedefaultseppunct}\relax
\EndOfBibitem
\bibitem[Kirkwood(1933)]{PhysRev.44.31}
Kirkwood,~J.~G. Quantum statistics of almost classical assemblies. \emph{Phys.
  Rev.} \textbf{1933}, \emph{44}, 31--37\relax
\mciteBstWouldAddEndPuncttrue
\mciteSetBstMidEndSepPunct{\mcitedefaultmidpunct}
{\mcitedefaultendpunct}{\mcitedefaultseppunct}\relax
\EndOfBibitem
\bibitem[Brack et~al.(1976)Brack, Jennings, and Chu]{brack76}
Brack,~M.; Jennings,~B.~K.; Chu,~Y.~H. On the extended Thomas--Fermi
  approximation to the kinetic energy density. \emph{Phys. Lett. B}
  \textbf{1976}, \emph{65}, 1--4\relax
\mciteBstWouldAddEndPuncttrue
\mciteSetBstMidEndSepPunct{\mcitedefaultmidpunct}
{\mcitedefaultendpunct}{\mcitedefaultseppunct}\relax
\EndOfBibitem
\bibitem[March(1977)]{march77}
March,~N.~H. Partial summation of gradient expansion of canonical density
  matrix. \emph{Phys. Lett. A} \textbf{1977}, \emph{64}, 185--186\relax
\mciteBstWouldAddEndPuncttrue
\mciteSetBstMidEndSepPunct{\mcitedefaultmidpunct}
{\mcitedefaultendpunct}{\mcitedefaultseppunct}\relax
\EndOfBibitem
\bibitem[Jennings(1978)]{jennings78}
Jennings,~B.~K. The extended Thomas--Fermi density matrix. \emph{Phys. Lett. B}
  \textbf{1978}, \emph{74}, 13--14\relax
\mciteBstWouldAddEndPuncttrue
\mciteSetBstMidEndSepPunct{\mcitedefaultmidpunct}
{\mcitedefaultendpunct}{\mcitedefaultseppunct}\relax
\EndOfBibitem
\bibitem[Engel and Dreizler(1989)Engel, and Dreizler]{engel89}
Engel,~E.; Dreizler,~R.~M. Extension of the
  Thomas--Fermi--Dirac--Weizs\"{a}cker model: four-order gradient corrections
  to the kinetic energy. \emph{J. Phys. B} \textbf{1989}, \emph{22}, 1901\relax
\mciteBstWouldAddEndPuncttrue
\mciteSetBstMidEndSepPunct{\mcitedefaultmidpunct}
{\mcitedefaultendpunct}{\mcitedefaultseppunct}\relax
\EndOfBibitem
\bibitem[Yang(1986)]{PhysRevA.34.4575}
Yang,~W. Gradient correction in Thomas--Fermi theory. \emph{Phys. Rev. A}
  \textbf{1986}, \emph{34}, 4575--4585\relax
\mciteBstWouldAddEndPuncttrue
\mciteSetBstMidEndSepPunct{\mcitedefaultmidpunct}
{\mcitedefaultendpunct}{\mcitedefaultseppunct}\relax
\EndOfBibitem
\bibitem[Scott(1952)]{Scott}
Scott,~J. The binding energy of the Thomas--Fermi Atom. \emph{Philos. Mag.}
  \textbf{1952}, \emph{43}, 859\relax
\mciteBstWouldAddEndPuncttrue
\mciteSetBstMidEndSepPunct{\mcitedefaultmidpunct}
{\mcitedefaultendpunct}{\mcitedefaultseppunct}\relax
\EndOfBibitem
\bibitem[Golden(1957)]{PhysRev.105.604}
Golden,~S. Statistical theory of many--electron systems. General considerations
  pertaining to the Thomas--Fermi theory. \emph{Phys. Rev.} \textbf{1957},
  \emph{105}, 604--615\relax
\mciteBstWouldAddEndPuncttrue
\mciteSetBstMidEndSepPunct{\mcitedefaultmidpunct}
{\mcitedefaultendpunct}{\mcitedefaultseppunct}\relax
\EndOfBibitem
\bibitem[Jones and Young(1971)Jones, and Young]{JY}
Jones,~W.; Young,~W.~H. Density functional theory and the von Weizs\"{a}cker
  method. \emph{J. Phys. C} \textbf{1971}, \emph{4}, 1322\relax
\mciteBstWouldAddEndPuncttrue
\mciteSetBstMidEndSepPunct{\mcitedefaultmidpunct}
{\mcitedefaultendpunct}{\mcitedefaultseppunct}\relax
\EndOfBibitem
\bibitem[Hodges(1973)]{Ho}
Hodges,~C.~H. Quantum corrections to the Thomas--Fermi approximation: the
  Kirzhnits method. \emph{Can. J. Phys.} \textbf{1973}, \emph{51}, 1428\relax
\mciteBstWouldAddEndPuncttrue
\mciteSetBstMidEndSepPunct{\mcitedefaultmidpunct}
{\mcitedefaultendpunct}{\mcitedefaultseppunct}\relax
\EndOfBibitem
\bibitem[Wang and Teter(1992)Wang, and Teter]{LWT92}
Wang,~L.-W.; Teter,~M.~P. Kinetic--energy functional of the electron density.
  \emph{Phys. Rev. B} \textbf{1992}, \emph{45}, 13196--13220\relax
\mciteBstWouldAddEndPuncttrue
\mciteSetBstMidEndSepPunct{\mcitedefaultmidpunct}
{\mcitedefaultendpunct}{\mcitedefaultseppunct}\relax
\EndOfBibitem
\bibitem[Levy and Ou-Yang(1988)Levy, and Ou-Yang]{pauli}
Levy,~M.; Ou-Yang,~H. Exact properties of the Pauli potential for the square
  root of the electron density and the kinetic energy functional. \emph{Phys.
  Rev. A} \textbf{1988}, \emph{38}, 625--629\relax
\mciteBstWouldAddEndPuncttrue
\mciteSetBstMidEndSepPunct{\mcitedefaultmidpunct}
{\mcitedefaultendpunct}{\mcitedefaultseppunct}\relax
\EndOfBibitem
\bibitem[Bartolotti and Acharya(1982)Bartolotti, and Acharya]{bartolotti:4576}
Bartolotti,~L.~J.; Acharya,~P.~K. On the functional derivative of the kinetic
  energy density functional. \emph{J. Chem. Phys.} \textbf{1982}, \emph{77},
  4576--4585\relax
\mciteBstWouldAddEndPuncttrue
\mciteSetBstMidEndSepPunct{\mcitedefaultmidpunct}
{\mcitedefaultendpunct}{\mcitedefaultseppunct}\relax
\EndOfBibitem
\bibitem[Yang et~al.(1986)Yang, Parr, and Lee]{PhysRevA.34.4586}
Yang,~W.; Parr,~R.~G.; Lee,~C. Various functionals for the kinetic energy
  density of an atom or molecule. \emph{Phys. Rev. A} \textbf{1986}, \emph{34},
  4586--4590\relax
\mciteBstWouldAddEndPuncttrue
\mciteSetBstMidEndSepPunct{\mcitedefaultmidpunct}
{\mcitedefaultendpunct}{\mcitedefaultseppunct}\relax
\EndOfBibitem
\bibitem[Karasiev et~al.(2000)Karasiev, Lude\~na, and Artemyev]{karasiev00}
Karasiev,~V.~V.; Lude\~na,~E.~V.; Artemyev,~A.~N. Electronic--structure
  kinetic--energy functional based on atomic local--scaling transformations.
  \emph{Phys. Rev. A} \textbf{2000}, \emph{62}, 062510\relax
\mciteBstWouldAddEndPuncttrue
\mciteSetBstMidEndSepPunct{\mcitedefaultmidpunct}
{\mcitedefaultendpunct}{\mcitedefaultseppunct}\relax
\EndOfBibitem
\bibitem[{Pis Diez} and Karasiev(2003){Pis Diez}, and Karasiev]{pisdies}
{Pis Diez},~R.; Karasiev,~V.~V. A relationship between the weighted density
  approximation and the local--scaling transformation version of density
  functional theory. \emph{J. Phys. B} \textbf{2003}, \emph{36}, 2881\relax
\mciteBstWouldAddEndPuncttrue
\mciteSetBstMidEndSepPunct{\mcitedefaultmidpunct}
{\mcitedefaultendpunct}{\mcitedefaultseppunct}\relax
\EndOfBibitem
\bibitem[Lee et~al.(1991)Lee, Lee, and Parr]{LLP91}
Lee,~H.; Lee,~C.; Parr,~R.~G. Conjoint gradient correction to the Hartree--Fock
  kinetic- and exchange--energy density functionals. \emph{Phys. Rev. A}
  \textbf{1991}, \emph{44}, 768--771\relax
\mciteBstWouldAddEndPuncttrue
\mciteSetBstMidEndSepPunct{\mcitedefaultmidpunct}
{\mcitedefaultendpunct}{\mcitedefaultseppunct}\relax
\EndOfBibitem
\bibitem[March and Santamaria(1991)March, and Santamaria]{QUA:QUA560390405}
March,~N.~H.; Santamaria,~R. Non--local relation between kinetic and exchange
  energy densities in Hartree--Fock theory. \emph{Int. J. Quant. Chem.}
  \textbf{1991}, \emph{39}, 585--592\relax
\mciteBstWouldAddEndPuncttrue
\mciteSetBstMidEndSepPunct{\mcitedefaultmidpunct}
{\mcitedefaultendpunct}{\mcitedefaultseppunct}\relax
\EndOfBibitem
\bibitem[Perdew(1992)]{perdewk92}
Perdew,~J.~P. Generalized gradient approximation for the fermion kinetic energy
  as a functional of the density. \emph{Phys. Lett. A} \textbf{1992},
  \emph{165}, 79--82\relax
\mciteBstWouldAddEndPuncttrue
\mciteSetBstMidEndSepPunct{\mcitedefaultmidpunct}
{\mcitedefaultendpunct}{\mcitedefaultseppunct}\relax
\EndOfBibitem
\bibitem[Lacks and Gordon(1994)Lacks, and Gordon]{lacks94}
Lacks,~D.~J.; Gordon,~R.~G. Tests of nonlocal kinetic energy functionals.
  \emph{J. Chem. Phys.} \textbf{1994}, \emph{100}, 4446--4452\relax
\mciteBstWouldAddEndPuncttrue
\mciteSetBstMidEndSepPunct{\mcitedefaultmidpunct}
{\mcitedefaultendpunct}{\mcitedefaultseppunct}\relax
\EndOfBibitem
\bibitem[DePristo and Kress(1987)DePristo, and Kress]{DK87}
DePristo,~A.~E.; Kress,~J.~D. Kinetic--energy functionals via Pad\'e
  approximations. \emph{Phys. Rev. A} \textbf{1987}, \emph{35}, 438--441\relax
\mciteBstWouldAddEndPuncttrue
\mciteSetBstMidEndSepPunct{\mcitedefaultmidpunct}
{\mcitedefaultendpunct}{\mcitedefaultseppunct}\relax
\EndOfBibitem
\bibitem[Thakkar(1992)]{Thak92}
Thakkar,~A.~J. Comparison of kinetic--energy density functionals. \emph{Phys.
  Rev. A} \textbf{1992}, \emph{46}, 6920--6924\relax
\mciteBstWouldAddEndPuncttrue
\mciteSetBstMidEndSepPunct{\mcitedefaultmidpunct}
{\mcitedefaultendpunct}{\mcitedefaultseppunct}\relax
\EndOfBibitem
\bibitem[Tran and Wesolowski(2002)Tran, and Wesolowski]{TW02}
Tran,~F.; Wesolowski,~T.~A. Link between the kinetic- and exchange--energy
  functionals in the generalized gradient approximation. \emph{Int. J. Quant.
  Chem.} \textbf{2002}, \emph{89}, 441\relax
\mciteBstWouldAddEndPuncttrue
\mciteSetBstMidEndSepPunct{\mcitedefaultmidpunct}
{\mcitedefaultendpunct}{\mcitedefaultseppunct}\relax
\EndOfBibitem
\bibitem[Karasiev et~al.(2006)Karasiev, Trickey, and Harris]{karasiev06}
Karasiev,~V.~V.; Trickey,~S.~B.; Harris,~F.~E. Born-Oppenheimer interatomic
  forces from simple, local kinetic energy density functionals. \emph{J. Comp.
  Aided Mat. Des.} \textbf{2006}, \emph{13}, 111\relax
\mciteBstWouldAddEndPuncttrue
\mciteSetBstMidEndSepPunct{\mcitedefaultmidpunct}
{\mcitedefaultendpunct}{\mcitedefaultseppunct}\relax
\EndOfBibitem
\bibitem[Karasiev et~al.(2009)Karasiev, Jones, Trickey, and
  Harris]{PhysRevB.80.245120}
Karasiev,~V.~V.; Jones,~R.~S.; Trickey,~S.~B.; Harris,~F.~E. Properties of
  constraint--based single--point approximate kinetic energy functionals.
  \emph{Phys. Rev. B} \textbf{2009}, \emph{80}, 245120\relax
\mciteBstWouldAddEndPuncttrue
\mciteSetBstMidEndSepPunct{\mcitedefaultmidpunct}
{\mcitedefaultendpunct}{\mcitedefaultseppunct}\relax
\EndOfBibitem
\bibitem[Huang and Carter(2010)Huang, and Carter]{huangcarter10}
Huang,~C.; Carter,~E.~A. Nonlocal orbital--free kinetic energy density
  functional for semiconductors. \emph{Phys. Rev. B} \textbf{2010}, \emph{81},
  045206\relax
\mciteBstWouldAddEndPuncttrue
\mciteSetBstMidEndSepPunct{\mcitedefaultmidpunct}
{\mcitedefaultendpunct}{\mcitedefaultseppunct}\relax
\EndOfBibitem
\bibitem[Wang et~al.(1998)Wang, Govind, and Carter]{WGC98}
Wang,~Y.~A.; Govind,~N.; Carter,~E.~A. Orbital--free kinetic--energy
  functionals for the nearly free electron gas. \emph{Phys. Rev. B}
  \textbf{1998}, \emph{58}, 13465--13471\relax
\mciteBstWouldAddEndPuncttrue
\mciteSetBstMidEndSepPunct{\mcitedefaultmidpunct}
{\mcitedefaultendpunct}{\mcitedefaultseppunct}\relax
\EndOfBibitem
\bibitem[Wang et~al.(2001)Wang, Govind, and Carter]{WGC_err}
Wang,~Y.~A.; Govind,~N.; Carter,~E.~A. Erratum: Orbital--free kinetic--energy
  functionals for the nearly free electron gas [Phys. Rev. B 58, 13 465
  (1998)]. \emph{Phys. Rev. B} \textbf{2001}, \emph{64}, 129901\relax
\mciteBstWouldAddEndPuncttrue
\mciteSetBstMidEndSepPunct{\mcitedefaultmidpunct}
{\mcitedefaultendpunct}{\mcitedefaultseppunct}\relax
\EndOfBibitem
\bibitem[Alonso and Girifalco(1978)Alonso, and Girifalco]{PhysRevB.17.3735}
Alonso,~J.~A.; Girifalco,~L.~A. Nonlocal approximation to the exchange
  potential and kinetic energy of an inhomogeneous electron gas. \emph{Phys.
  Rev. B} \textbf{1978}, \emph{17}, 3735--3743\relax
\mciteBstWouldAddEndPuncttrue
\mciteSetBstMidEndSepPunct{\mcitedefaultmidpunct}
{\mcitedefaultendpunct}{\mcitedefaultseppunct}\relax
\EndOfBibitem
\bibitem[Chac\'on et~al.(1985)Chac\'on, Alvarellos, and
  Tarazona]{PhysRevB.32.7868}
Chac\'on,~E.; Alvarellos,~J.~E.; Tarazona,~P. Nonlocal kinetic energy
  functional for nonhomogeneous electron systems. \emph{Phys. Rev. B}
  \textbf{1985}, \emph{32}, 7868--7877\relax
\mciteBstWouldAddEndPuncttrue
\mciteSetBstMidEndSepPunct{\mcitedefaultmidpunct}
{\mcitedefaultendpunct}{\mcitedefaultseppunct}\relax
\EndOfBibitem
\bibitem[Garc\'ia-Gonz\'alez et~al.(1996)Garc\'ia-Gonz\'alez, Alvarellos, and
  Chac\'on]{PhysRevB.53.9509}
Garc\'ia-Gonz\'alez,~P.; Alvarellos,~J.~E.; Chac\'on,~E. Nonlocal
  kinetic--energy--density functionals. \emph{Phys. Rev. B} \textbf{1996},
  \emph{53}, 9509--9512\relax
\mciteBstWouldAddEndPuncttrue
\mciteSetBstMidEndSepPunct{\mcitedefaultmidpunct}
{\mcitedefaultendpunct}{\mcitedefaultseppunct}\relax
\EndOfBibitem
\bibitem[Garc\'ia-Gonz\'alez et~al.(1996)Garc\'ia-Gonz\'alez, Alvarellos, and
  Chac\'on]{PhysRevA.54.1897}
Garc\'ia-Gonz\'alez,~P.; Alvarellos,~J.~E.; Chac\'on,~E. Kinetic--energy
  density functional: Atoms and shell structure. \emph{Phys. Rev. A}
  \textbf{1996}, \emph{54}, 1897--1905\relax
\mciteBstWouldAddEndPuncttrue
\mciteSetBstMidEndSepPunct{\mcitedefaultmidpunct}
{\mcitedefaultendpunct}{\mcitedefaultseppunct}\relax
\EndOfBibitem
\bibitem[Garc\'ia-Gonz\'alez et~al.(1998)Garc\'ia-Gonz\'alez, Alvarellos, and
  Chac\'on]{PhysRevB.57.4857}
Garc\'ia-Gonz\'alez,~P.; Alvarellos,~J.~E.; Chac\'on,~E. Nonlocal symmetrized
  kinetic--energy density functional: Application to simple surfaces.
  \emph{Phys. Rev. B} \textbf{1998}, \emph{57}, 4857--4862\relax
\mciteBstWouldAddEndPuncttrue
\mciteSetBstMidEndSepPunct{\mcitedefaultmidpunct}
{\mcitedefaultendpunct}{\mcitedefaultseppunct}\relax
\EndOfBibitem
\bibitem[Garc\'ia-Gonz\'alez et~al.(1998)Garc\'ia-Gonz\'alez, Alvarellos, and
  Chac\'on]{PhysRevA.57.4192}
Garc\'ia-Gonz\'alez,~P.; Alvarellos,~J.~E.; Chac\'on,~E. Kinetic--energy
  density functionals based on the homogeneous response function applied to
  one--dimensional fermion systems. \emph{Phys. Rev. A} \textbf{1998},
  \emph{57}, 4192--4200\relax
\mciteBstWouldAddEndPuncttrue
\mciteSetBstMidEndSepPunct{\mcitedefaultmidpunct}
{\mcitedefaultendpunct}{\mcitedefaultseppunct}\relax
\EndOfBibitem
\bibitem[Smargiassi and Madden(1994)Smargiassi, and Madden]{PhysRevB.49.5220}
Smargiassi,~E.; Madden,~P.~A. Orbital--free kinetic-energy functionals for
  first--principles molecular dynamics. \emph{Phys. Rev. B} \textbf{1994},
  \emph{49}, 5220--5226\relax
\mciteBstWouldAddEndPuncttrue
\mciteSetBstMidEndSepPunct{\mcitedefaultmidpunct}
{\mcitedefaultendpunct}{\mcitedefaultseppunct}\relax
\EndOfBibitem
\bibitem[Foley and Madden(1996)Foley, and Madden]{PhysRevB.53.10589}
Foley,~M.; Madden,~P.~A. Further orbital--free kinetic-energy functionals for
  ab initio molecular dynamics. \emph{Phys. Rev. B} \textbf{1996}, \emph{53},
  10589--10598\relax
\mciteBstWouldAddEndPuncttrue
\mciteSetBstMidEndSepPunct{\mcitedefaultmidpunct}
{\mcitedefaultendpunct}{\mcitedefaultseppunct}\relax
\EndOfBibitem
\bibitem[Wang et~al.(1999)Wang, Govind, and Carter]{PhysRevB.60.16350}
Wang,~Y.~A.; Govind,~N.; Carter,~E.~A. Orbital--free kinetic--energy density
  functionals with a density--dependent kernel. \emph{Phys. Rev. B}
  \textbf{1999}, \emph{60}, 16350--16358\relax
\mciteBstWouldAddEndPuncttrue
\mciteSetBstMidEndSepPunct{\mcitedefaultmidpunct}
{\mcitedefaultendpunct}{\mcitedefaultseppunct}\relax
\EndOfBibitem
\bibitem[Karasiev et~al.(1989)Karasiev, Jones, Trickey, and
  Harris]{karasiev_chap}
Karasiev,~V.~V.; Jones,~R.~S.; Trickey,~S.~B.; Harris,~F.~E. \emph{New
  Developments in Quantum Chemistry}; Transworld Research Network:
  Trivandrum-695 023, Kerala, India, 1989; Chapter Recent advances in
  developing orbital--free kinetic energy functionals\relax
\mciteBstWouldAddEndPuncttrue
\mciteSetBstMidEndSepPunct{\mcitedefaultmidpunct}
{\mcitedefaultendpunct}{\mcitedefaultseppunct}\relax
\EndOfBibitem
\bibitem[Acharya et~al.(1980)Acharya, Bartolotti, Sears, and
  Parr]{Acharya01121980}
Acharya,~P.~K.; Bartolotti,~L.~J.; Sears,~S.~B.; Parr,~R.~G. An atomic kinetic
  energy functional with full Weizsacker correction. \emph{Proc. Nat. Acad.
  Sci.} \textbf{1980}, \emph{77}, 6978--6982\relax
\mciteBstWouldAddEndPuncttrue
\mciteSetBstMidEndSepPunct{\mcitedefaultmidpunct}
{\mcitedefaultendpunct}{\mcitedefaultseppunct}\relax
\EndOfBibitem
\bibitem[Ou-Yang and Levy(1991)Ou-Yang, and Levy]{OL91}
Ou-Yang,~H.; Levy,~M. Approximate noninteracting kinetic energy functionals
  from a nonuniform scaling requirement. \emph{Int. J. Quant. Chem.}
  \textbf{1991}, \emph{40}, 379\relax
\mciteBstWouldAddEndPuncttrue
\mciteSetBstMidEndSepPunct{\mcitedefaultmidpunct}
{\mcitedefaultendpunct}{\mcitedefaultseppunct}\relax
\EndOfBibitem
\bibitem[Vitos et~al.(1998)Vitos, Skriver, and Koll\'ar]{vitosjellium98}
Vitos,~L.; Skriver,~H.~L.; Koll\'ar,~J. Kinetic--energy functionals studied by
  surface calculations. \emph{Phys. Rev. B} \textbf{1998}, \emph{57},
  12611--12615\relax
\mciteBstWouldAddEndPuncttrue
\mciteSetBstMidEndSepPunct{\mcitedefaultmidpunct}
{\mcitedefaultendpunct}{\mcitedefaultseppunct}\relax
\EndOfBibitem
\bibitem[Perdew and Constantin(2007)Perdew, and Constantin]{PhysRevB.75.155109}
Perdew,~J.~P.; Constantin,~L.~A. Laplacian--level density functionals for the
  kinetic energy density and exchange--correlation energy. \emph{Phys. Rev. B}
  \textbf{2007}, \emph{75}, 155109\relax
\mciteBstWouldAddEndPuncttrue
\mciteSetBstMidEndSepPunct{\mcitedefaultmidpunct}
{\mcitedefaultendpunct}{\mcitedefaultseppunct}\relax
\EndOfBibitem
\bibitem[Constantin and Ruzsinszky(2009)Constantin, and
  Ruzsinszky]{PhysRevB.79.115117}
Constantin,~L.~A.; Ruzsinszky,~A. Kinetic energy density functionals from the
  Airy gas with an application to the atomization kinetic energies of
  molecules. \emph{Phys. Rev. B} \textbf{2009}, \emph{79}, 115117\relax
\mciteBstWouldAddEndPuncttrue
\mciteSetBstMidEndSepPunct{\mcitedefaultmidpunct}
{\mcitedefaultendpunct}{\mcitedefaultseppunct}\relax
\EndOfBibitem
\bibitem[Garc\'{\i}a-Aldea and Alvarellos(2007)Garc\'{\i}a-Aldea, and
  Alvarellos]{alvarellos07}
Garc\'{\i}a-Aldea,~D.; Alvarellos,~J.~E. Kinetic energy density study of some
  representative semilocal kinetic energy functionals. \emph{J. Chem. Phys.}
  \textbf{2007}, \emph{127}, 144109\relax
\mciteBstWouldAddEndPuncttrue
\mciteSetBstMidEndSepPunct{\mcitedefaultmidpunct}
{\mcitedefaultendpunct}{\mcitedefaultseppunct}\relax
\EndOfBibitem
\bibitem[Garc\'{\i}a-Aldea and Alvarellos(2008)Garc\'{\i}a-Aldea, and
  Alvarellos]{alvarellos08}
Garc\'{\i}a-Aldea,~D.; Alvarellos,~J.~E. Fully nonlocal kinetic energy density
  functionals: A proposal and a general assessment for atomic systems. \emph{J.
  Chem. Phys.} \textbf{2008}, \emph{129}, 074103\relax
\mciteBstWouldAddEndPuncttrue
\mciteSetBstMidEndSepPunct{\mcitedefaultmidpunct}
{\mcitedefaultendpunct}{\mcitedefaultseppunct}\relax
\EndOfBibitem
\bibitem[Garc\'\i{}a-Aldea and Alvarellos(2008)Garc\'\i{}a-Aldea, and
  Alvarellos]{alvarellos08b}
Garc\'\i{}a-Aldea,~D.; Alvarellos,~J.~E. Approach to kinetic energy density
  functionals: Nonlocal terms with the structure of the von Weizs\"acker
  functional. \emph{Phys. Rev. A} \textbf{2008}, \emph{77}, 022502\relax
\mciteBstWouldAddEndPuncttrue
\mciteSetBstMidEndSepPunct{\mcitedefaultmidpunct}
{\mcitedefaultendpunct}{\mcitedefaultseppunct}\relax
\EndOfBibitem
\bibitem[Chai and Weeks(2007)Chai, and Weeks]{PhysRevB.75.205122}
Chai,~J.-D.; Weeks,~J.~D. Orbital--free density functional theory: Kinetic
  potentials and ab initio local pseudopotentials. \emph{Phys. Rev. B}
  \textbf{2007}, \emph{75}, 205122\relax
\mciteBstWouldAddEndPuncttrue
\mciteSetBstMidEndSepPunct{\mcitedefaultmidpunct}
{\mcitedefaultendpunct}{\mcitedefaultseppunct}\relax
\EndOfBibitem
\bibitem[Lembarki and Chermette(1994)Lembarki, and Chermette]{LC94}
Lembarki,~A.; Chermette,~H. Obtaining a gradient--corrected kinetic--energy
  functional from the Perdew--Wang exchange functional. \emph{Phys. Rev. A}
  \textbf{1994}, \emph{50}, 5328\relax
\mciteBstWouldAddEndPuncttrue
\mciteSetBstMidEndSepPunct{\mcitedefaultmidpunct}
{\mcitedefaultendpunct}{\mcitedefaultseppunct}\relax
\EndOfBibitem
\bibitem[G\"otz et~al.(2009)G\"otz, Beyhan, and Visscher]{gotz09}
G\"otz,~A.~W.; Beyhan,~S.~M.; Visscher,~L. Performance of kinetic energy
  functionals for interaction energies in a subsystem formulation of density
  functional theory. \emph{J. Chem. Theory Comput.} \textbf{2009}, \emph{5},
  3161--3174\relax
\mciteBstWouldAddEndPuncttrue
\mciteSetBstMidEndSepPunct{\mcitedefaultmidpunct}
{\mcitedefaultendpunct}{\mcitedefaultseppunct}\relax
\EndOfBibitem
\bibitem[Constantin et~al.(2011)Constantin, Fabiano, Laricchia, and {Della
  Sala}]{apbek}
Constantin,~L.~A.; Fabiano,~E.; Laricchia,~S.; {Della Sala},~F. Semiclassical
  neutral atom as a reference system in density functional theory. \emph{Phys.
  Rev. Lett.} \textbf{2011}, \emph{106}, 186406\relax
\mciteBstWouldAddEndPuncttrue
\mciteSetBstMidEndSepPunct{\mcitedefaultmidpunct}
{\mcitedefaultendpunct}{\mcitedefaultseppunct}\relax
\EndOfBibitem
\bibitem[Laricchia et~al.(2011)Laricchia, Fabiano, Constantin, and {Della
  Sala}]{apbekint}
Laricchia,~S.; Fabiano,~E.; Constantin,~L.~A.; {Della Sala},~F. Generalized
  gradient approximations of the noninteracting kinetic energy from the
  semiclassical atom theory: rationalization of the accuracy of the frozen
  density embedding theory for nonbonded interactions. \emph{J. Chem. Theory
  Comput.} \textbf{2011}, \emph{7}, 2439--2451\relax
\mciteBstWouldAddEndPuncttrue
\mciteSetBstMidEndSepPunct{\mcitedefaultmidpunct}
{\mcitedefaultendpunct}{\mcitedefaultseppunct}\relax
\EndOfBibitem
\bibitem[Tran and Wesolowski(2013)Tran, and Wesolowski]{weso_chap_funct}
Tran,~F.; Wesolowski,~T.~A. In \emph{Recent Advances in Computational Chemistry
  6}; Wesolowski,~T.~A., Wang,~Y.~A., Eds.; World Scientific: Singapore, 2013;
  pp 429--442\relax
\mciteBstWouldAddEndPuncttrue
\mciteSetBstMidEndSepPunct{\mcitedefaultmidpunct}
{\mcitedefaultendpunct}{\mcitedefaultseppunct}\relax
\EndOfBibitem
\bibitem[Tran and Wesolowski(2002)Tran, and Wesolowski]{TWESOm}
Tran,~F.; Wesolowski,~T.~A. Introduction of the explicit long--range
  nonlocality as an alternative to the gradient expansion approximation for the
  kinetic--energy functional. \emph{Chem. Phys. Lett.} \textbf{2002},
  \emph{360}, 209 -- 216\relax
\mciteBstWouldAddEndPuncttrue
\mciteSetBstMidEndSepPunct{\mcitedefaultmidpunct}
{\mcitedefaultendpunct}{\mcitedefaultseppunct}\relax
\EndOfBibitem
\bibitem[Pearson and Gordon(1985)Pearson, and Gordon]{PeGor}
Pearson,~E.~W.; Gordon,~R.~G. Local asymptotic gradient corrections to the
  energy functional of an electron gas. \emph{J. Chem. Phys.} \textbf{1985},
  \emph{82}, 881--889\relax
\mciteBstWouldAddEndPuncttrue
\mciteSetBstMidEndSepPunct{\mcitedefaultmidpunct}
{\mcitedefaultendpunct}{\mcitedefaultseppunct}\relax
\EndOfBibitem
\bibitem[Allan et~al.(1985)Allan, West, Cooper, Grout, and March]{allan85}
Allan,~N.~L.; West,~C.~G.; Cooper,~D.~L.; Grout,~P.~J.; March,~N.~H. The
  gradient expansions of the kinetic energy and the mean momentum for light
  diatomic molecules. \emph{J. Chem. Phys.} \textbf{1985}, \emph{83},
  4562--4564\relax
\mciteBstWouldAddEndPuncttrue
\mciteSetBstMidEndSepPunct{\mcitedefaultmidpunct}
{\mcitedefaultendpunct}{\mcitedefaultseppunct}\relax
\EndOfBibitem
\bibitem[Huang and Carter(2006)Huang, and Carter]{huang06}
Huang,~P.; Carter,~E.~A. Self-consistent embedding theory for locally
  correlated configuration interaction wave functions in condensed matter.
  \emph{J. Chem. Phys.} \textbf{2006}, \emph{125}, 084102\relax
\mciteBstWouldAddEndPuncttrue
\mciteSetBstMidEndSepPunct{\mcitedefaultmidpunct}
{\mcitedefaultendpunct}{\mcitedefaultseppunct}\relax
\EndOfBibitem
\bibitem[Wang and Carter(2000)Wang, and Carter]{wangcarterof}
Wang,~Y.; Carter,~E.~A. In \emph{Progress in Theoretical Chemistry and
  Physics}; Schwartz,~S., Ed.; Kluwer: Dordrecht, 2000; p 117\relax
\mciteBstWouldAddEndPuncttrue
\mciteSetBstMidEndSepPunct{\mcitedefaultmidpunct}
{\mcitedefaultendpunct}{\mcitedefaultseppunct}\relax
\EndOfBibitem
\bibitem[Watson and Carter(2000)Watson, and Carter]{Watson00}
Watson,~S.~C.; Carter,~E.~A. Linear-scaling parallel algorithms for the first
  principles treatment of metals. \emph{Comput. Phys. Commun.} \textbf{2000},
  \emph{128}, 67 -- 92\relax
\mciteBstWouldAddEndPuncttrue
\mciteSetBstMidEndSepPunct{\mcitedefaultmidpunct}
{\mcitedefaultendpunct}{\mcitedefaultseppunct}\relax
\EndOfBibitem
\bibitem[Govind et~al.(1994)Govind, Wang, and Guo]{govind94}
Govind,~N.; Wang,~J.; Guo,~H. Total-energy calculations using a
  gradient-expanded kinetic-energy functional. \emph{Phys. Rev. B}
  \textbf{1994}, \emph{50}, 11175--11178\relax
\mciteBstWouldAddEndPuncttrue
\mciteSetBstMidEndSepPunct{\mcitedefaultmidpunct}
{\mcitedefaultendpunct}{\mcitedefaultseppunct}\relax
\EndOfBibitem
\bibitem[Zhou et~al.(2005)Zhou, Ligneres, and Carter]{zhou05}
Zhou,~B.; Ligneres,~V.~L.; Carter,~E.~A. Improving the orbital-free density
  functional theory description of covalent materials. \emph{J. Chem. Phys.}
  \textbf{2005}, \emph{122}, 044103\relax
\mciteBstWouldAddEndPuncttrue
\mciteSetBstMidEndSepPunct{\mcitedefaultmidpunct}
{\mcitedefaultendpunct}{\mcitedefaultseppunct}\relax
\EndOfBibitem
\bibitem[Smargiassi and Madden(1994)Smargiassi, and Madden]{madden94}
Smargiassi,~E.; Madden,~P.~A. Orbital-free kinetic-energy functionals for
  first-principles molecular dynamics. \emph{Phys. Rev. B} \textbf{1994},
  \emph{49}, 5220--5226\relax
\mciteBstWouldAddEndPuncttrue
\mciteSetBstMidEndSepPunct{\mcitedefaultmidpunct}
{\mcitedefaultendpunct}{\mcitedefaultseppunct}\relax
\EndOfBibitem
\bibitem[Pearson et~al.(1993)Pearson, Smargiassi, and Madden]{madden93}
Pearson,~M.; Smargiassi,~E.; Madden,~P.~A. Ab initio molecular dynamics with an
  orbital-free density functional. \emph{J. Phys. Cond. Matt.} \textbf{1993},
  \emph{5}, 3221\relax
\mciteBstWouldAddEndPuncttrue
\mciteSetBstMidEndSepPunct{\mcitedefaultmidpunct}
{\mcitedefaultendpunct}{\mcitedefaultseppunct}\relax
\EndOfBibitem
\bibitem[Foley and Madden(1996)Foley, and Madden]{madden96}
Foley,~M.; Madden,~P.~A. Further orbital-free kinetic-energy functionals for ab
  initio molecular dynamics. \emph{Phys. Rev. B} \textbf{1996}, \emph{53},
  10589--10598\relax
\mciteBstWouldAddEndPuncttrue
\mciteSetBstMidEndSepPunct{\mcitedefaultmidpunct}
{\mcitedefaultendpunct}{\mcitedefaultseppunct}\relax
\EndOfBibitem
\bibitem[Xia and Carter(2012)Xia, and Carter]{CarterXia}
Xia,~J.; Carter,~E.~A. Density-decomposed orbital-free density functional
  theory for covalently bonded molecules and materials. \emph{Phys. Rev. B}
  \textbf{2012}, \emph{86}, 235109\relax
\mciteBstWouldAddEndPuncttrue
\mciteSetBstMidEndSepPunct{\mcitedefaultmidpunct}
{\mcitedefaultendpunct}{\mcitedefaultseppunct}\relax
\EndOfBibitem
\bibitem[Huang and Carter(2012)Huang, and Carter]{PhysRevB.85.045126}
Huang,~C.; Carter,~E.~A. Toward an orbital-free density functional theory of
  transition metals based on an electron density decomposition. \emph{Phys.
  Rev. B} \textbf{2012}, \emph{85}, 045126\relax
\mciteBstWouldAddEndPuncttrue
\mciteSetBstMidEndSepPunct{\mcitedefaultmidpunct}
{\mcitedefaultendpunct}{\mcitedefaultseppunct}\relax
\EndOfBibitem
\bibitem[Snyder et~al.(2012)Snyder, Rupp, Hansen, M\"{u}ller, and
  Burke]{Kieronml}
Snyder,~J.; Rupp,~M.; Hansen,~K.; M\"{u}ller,~K.-R.; Burke,~K. Finding density
  functionals with machine learning. \emph{Phys. Rev. Lett.} \textbf{2012},
  \emph{108}, 253002\relax
\mciteBstWouldAddEndPuncttrue
\mciteSetBstMidEndSepPunct{\mcitedefaultmidpunct}
{\mcitedefaultendpunct}{\mcitedefaultseppunct}\relax
\EndOfBibitem
\bibitem[Gordon and Kim(1972)Gordon, and Kim]{GoKim}
Gordon,~R.~G.; Kim,~Y.~S. Theory for the forces between closed-shell atoms and
  molecules. \emph{J. Chem. Phys.} \textbf{1972}, \emph{56}, 3122--3133\relax
\mciteBstWouldAddEndPuncttrue
\mciteSetBstMidEndSepPunct{\mcitedefaultmidpunct}
{\mcitedefaultendpunct}{\mcitedefaultseppunct}\relax
\EndOfBibitem
\bibitem[Senatore and Subbaswamy(1986)Senatore, and Subbaswamy]{senatore86}
Senatore,~G.; Subbaswamy,~K.~R. Density dependence of the dielectric constant
  of rare-gas crystals. \emph{Phys. Rev. B} \textbf{1986}, \emph{34},
  5754--5757\relax
\mciteBstWouldAddEndPuncttrue
\mciteSetBstMidEndSepPunct{\mcitedefaultmidpunct}
{\mcitedefaultendpunct}{\mcitedefaultseppunct}\relax
\EndOfBibitem
\bibitem[Cortona(1991)]{cortona}
Cortona,~P. Self-consistently determined properties of solids without
  band-structure calculations. \emph{Phys. Rev. B} \textbf{1991}, \emph{44},
  8454\relax
\mciteBstWouldAddEndPuncttrue
\mciteSetBstMidEndSepPunct{\mcitedefaultmidpunct}
{\mcitedefaultendpunct}{\mcitedefaultseppunct}\relax
\EndOfBibitem
\bibitem[Wesolowski and Warshel(1993)Wesolowski, and Warshel]{wesowarh93}
Wesolowski,~T.~A.; Warshel,~A. Frozen density functional approach for ab initio
  calculations of solvated molecules. \emph{J. Phys. Chem.} \textbf{1993},
  \emph{97}, 8050\relax
\mciteBstWouldAddEndPuncttrue
\mciteSetBstMidEndSepPunct{\mcitedefaultmidpunct}
{\mcitedefaultendpunct}{\mcitedefaultseppunct}\relax
\EndOfBibitem
\bibitem[Wesolowski(2006)]{wesorev}
Wesolowski,~T.~A. In \emph{Chemistry: Reviews of Current Trends};
  Leszczynski,~J., Ed.; World Scientific: Singapore, 2006: Singapore, 2006;
  Vol.~10; p~1\relax
\mciteBstWouldAddEndPuncttrue
\mciteSetBstMidEndSepPunct{\mcitedefaultmidpunct}
{\mcitedefaultendpunct}{\mcitedefaultseppunct}\relax
\EndOfBibitem
\bibitem[Wesolowski et~al.(1996)Wesolowski, Chermette, and
  Weber]{wesolowski96fhnch}
Wesolowski,~T.~A.; Chermette,~H.; Weber,~J. Accuracy of approximate kinetic
  energy functionals in the model of Kohn--Sham equations with constrained
  electron density: The FH--NCH complex as a test case. \emph{J. Chem. Phys.}
  \textbf{1996}, \emph{105}, 9182--9190\relax
\mciteBstWouldAddEndPuncttrue
\mciteSetBstMidEndSepPunct{\mcitedefaultmidpunct}
{\mcitedefaultendpunct}{\mcitedefaultseppunct}\relax
\EndOfBibitem
\bibitem[Hodak et~al.(2008)Hodak, Lu, and Bernholc]{hodak08}
Hodak,~M.; Lu,~W.; Bernholc,~J. Hybrid ab initio Kohn--Sham density functional
  theory/frozen-density orbital-free density functional theory simulation
  method suitable for biological systems. \emph{J. Chem. Phys.} \textbf{2008},
  \emph{128}, 014101\relax
\mciteBstWouldAddEndPuncttrue
\mciteSetBstMidEndSepPunct{\mcitedefaultmidpunct}
{\mcitedefaultendpunct}{\mcitedefaultseppunct}\relax
\EndOfBibitem
\bibitem[Elliott et~al.(2009)Elliott, Cohen, Wasserman, and
  Burke]{elliot09jctc}
Elliott,~P.; Cohen,~M.~H.; Wasserman,~A.; Burke,~K. Density functional
  partition theory with fractional occupations. \emph{J. Chem. Theory Comput.}
  \textbf{2009}, \emph{5}, 827\relax
\mciteBstWouldAddEndPuncttrue
\mciteSetBstMidEndSepPunct{\mcitedefaultmidpunct}
{\mcitedefaultendpunct}{\mcitedefaultseppunct}\relax
\EndOfBibitem
\bibitem[Neugebauer(2010)]{neug10}
Neugebauer,~J. Chromophore-specific theoretical spectroscopy: From subsystem
  density functional theory to mode-specific vibrational spectroscopy.
  \emph{Phys. Rep.} \textbf{2010}, \emph{489}, 1\relax
\mciteBstWouldAddEndPuncttrue
\mciteSetBstMidEndSepPunct{\mcitedefaultmidpunct}
{\mcitedefaultendpunct}{\mcitedefaultseppunct}\relax
\EndOfBibitem
\bibitem[Laricchia et~al.(2010)Laricchia, Fabiano, and {Della
  Sala}]{laricchia10}
Laricchia,~S.; Fabiano,~E.; {Della Sala},~F. Frozen density embedding with
  hybrid functionals. \emph{J. Chem. Phys.} \textbf{2010}, \emph{133},
  164111\relax
\mciteBstWouldAddEndPuncttrue
\mciteSetBstMidEndSepPunct{\mcitedefaultmidpunct}
{\mcitedefaultendpunct}{\mcitedefaultseppunct}\relax
\EndOfBibitem
\bibitem[Laricchia et~al.(2011)Laricchia, Fabiano, and {Della
  Sala}]{Laricchia2011114}
Laricchia,~S.; Fabiano,~E.; {Della Sala},~F. Frozen density embedding
  calculations with the orbital--dependent localized Hartree-Fock Kohn-Sham
  potential. \emph{Chem. Phys. Lett.} \textbf{2011}, \emph{518}, 114 --
  118\relax
\mciteBstWouldAddEndPuncttrue
\mciteSetBstMidEndSepPunct{\mcitedefaultmidpunct}
{\mcitedefaultendpunct}{\mcitedefaultseppunct}\relax
\EndOfBibitem
\bibitem[Goodpaster et~al.(2010)Goodpaster, Ananth, Manby, and {Miller
  III}]{goodpaster10}
Goodpaster,~J.~D.; Ananth,~N.; Manby,~F.~R.; {Miller III},~T.~F. Exact
  nonadditive kinetic potentials for embedded density functional theory.
  \emph{J. Chem. Phys.} \textbf{2010}, \emph{133}, 084103\relax
\mciteBstWouldAddEndPuncttrue
\mciteSetBstMidEndSepPunct{\mcitedefaultmidpunct}
{\mcitedefaultendpunct}{\mcitedefaultseppunct}\relax
\EndOfBibitem
\bibitem[Wesolowski and Warshel(1996)Wesolowski, and Warshel]{wesowarshel96}
Wesolowski,~T.~A.; Warshel,~A. Kohn-Sham equations with constrained electron
  density: an iterative evaluation of the ground-state electron density of
  interacting molecules. \emph{Chem. Phys. Lett.} \textbf{1996}, \emph{248},
  71\relax
\mciteBstWouldAddEndPuncttrue
\mciteSetBstMidEndSepPunct{\mcitedefaultmidpunct}
{\mcitedefaultendpunct}{\mcitedefaultseppunct}\relax
\EndOfBibitem
\bibitem[Laricchia et~al.(2012)Laricchia, Fabiano, and {Della
  Sala}]{laricchia:014102}
Laricchia,~S.; Fabiano,~E.; {Della Sala},~F. On the accuracy of frozen density
  embedding calculations with hybrid and orbital-dependent functionals for
  non-bonded interaction energies. \emph{J. Chem. Phys.} \textbf{2012},
  \emph{137}, 014102\relax
\mciteBstWouldAddEndPuncttrue
\mciteSetBstMidEndSepPunct{\mcitedefaultmidpunct}
{\mcitedefaultendpunct}{\mcitedefaultseppunct}\relax
\EndOfBibitem
\bibitem[Laricchia et~al.(2013)Laricchia, Fabiano, and {Della
  Sala}]{laricchia:124112}
Laricchia,~S.; Fabiano,~E.; {Della Sala},~F. Semilocal and hybrid density
  embedding calculations of ground-state charge-transfer complexes. \emph{J.
  Chem. Phys.} \textbf{2013}, \emph{138}, 124112\relax
\mciteBstWouldAddEndPuncttrue
\mciteSetBstMidEndSepPunct{\mcitedefaultmidpunct}
{\mcitedefaultendpunct}{\mcitedefaultseppunct}\relax
\EndOfBibitem
\bibitem[Ernzerhof(2000)]{Ernzerhof200059}
Ernzerhof,~M. The role of the kinetic energy density in approximations to the
  exchange energy. \emph{J. Mol. Struct.: THEOCHEM} \textbf{2000},
  \emph{501-502}, 59 -- 64\relax
\mciteBstWouldAddEndPuncttrue
\mciteSetBstMidEndSepPunct{\mcitedefaultmidpunct}
{\mcitedefaultendpunct}{\mcitedefaultseppunct}\relax
\EndOfBibitem
\bibitem[Perdew et~al.(1999)Perdew, Kurth, Zupan, and Blaha]{PKZB}
Perdew,~J.~P.; Kurth,~S.; Zupan,~A.; Blaha,~P. Accurate density functional with
  correct formal properties: A step beyond the generalized gradient
  approximation. \emph{Phys. Rev. Lett.} \textbf{1999}, \emph{82}, 2544\relax
\mciteBstWouldAddEndPuncttrue
\mciteSetBstMidEndSepPunct{\mcitedefaultmidpunct}
{\mcitedefaultendpunct}{\mcitedefaultseppunct}\relax
\EndOfBibitem
\bibitem[Tao et~al.(2003)Tao, Perdew, Staroverov, and Scuseria]{tpss}
Tao,~J.; Perdew,~J.~P.; Staroverov,~V.~N.; Scuseria,~G.~E. Climbing the density
  functional ladder: Nonempirical meta--generalized gradient approximation
  designed for molecules and solids. \emph{Phys. Rev. Lett.} \textbf{2003},
  \emph{91}, 146401\relax
\mciteBstWouldAddEndPuncttrue
\mciteSetBstMidEndSepPunct{\mcitedefaultmidpunct}
{\mcitedefaultendpunct}{\mcitedefaultseppunct}\relax
\EndOfBibitem
\bibitem[Yang et~al.(1986)Yang, Parr, and Lee]{YPL}
Yang,~W.; Parr,~R.~G.; Lee,~C. Various functionals for the kinetic energy
  density of an atom or molecule. \emph{Phys. Rev. A} \textbf{1986}, \emph{34},
  4586\relax
\mciteBstWouldAddEndPuncttrue
\mciteSetBstMidEndSepPunct{\mcitedefaultmidpunct}
{\mcitedefaultendpunct}{\mcitedefaultseppunct}\relax
\EndOfBibitem
\bibitem[Hoffmann-Ostenhof and Hoffmann-Ostenhof(1978)Hoffmann-Ostenhof, and
  Hoffmann-Ostenhof]{HOHO}
Hoffmann-Ostenhof,~T.; Hoffmann-Ostenhof,~M. Bounds to expectation values and
  exponentially decreasing upper bounds to the one-electron density of atoms.
  \emph{J. Phys. B} \textbf{1978}, \emph{11}, 17\relax
\mciteBstWouldAddEndPuncttrue
\mciteSetBstMidEndSepPunct{\mcitedefaultmidpunct}
{\mcitedefaultendpunct}{\mcitedefaultseppunct}\relax
\EndOfBibitem
\bibitem[Lee et~al.(2009)Lee, Constantin, Perdew, and Burke]{LCPB09}
Lee,~D.; Constantin,~L.~A.; Perdew,~J.~P.; Burke,~K. Condition on the
  Kohn--Sham kinetic energy and modern parametrization of the Thomas--Fermi
  density. \emph{J. Chem. Phys.} \textbf{2009}, \emph{130}, 034107\relax
\mciteBstWouldAddEndPuncttrue
\mciteSetBstMidEndSepPunct{\mcitedefaultmidpunct}
{\mcitedefaultendpunct}{\mcitedefaultseppunct}\relax
\EndOfBibitem

\bibitem[Fabiano and Constantin(2013)Fabiano, and Constantin]{scaling}
Fabiano,~E.; Constantin,~L.~A. Relevance of coordinate and particle-number
  scaling in density-functional theory. \emph{Phys. Rev. A} \textbf{2013},
  \emph{87}, 012511\relax
\mciteBstWouldAddEndPuncttrue
\mciteSetBstMidEndSepPunct{\mcitedefaultmidpunct}
{\mcitedefaultendpunct}{\mcitedefaultseppunct}\relax
\EndOfBibitem
\bibitem[Heilmann and Lieb(1995)Heilmann, and Lieb]{Lieb1}
Heilmann,~O.~J.; Lieb,~E.~H. Electron density near the nucleus of a large atom.
  \emph{Phys. Rev. A} \textbf{1995}, \emph{52}, 3628\relax
\mciteBstWouldAddEndPuncttrue
\mciteSetBstMidEndSepPunct{\mcitedefaultmidpunct}
{\mcitedefaultendpunct}{\mcitedefaultseppunct}\relax
\EndOfBibitem
\bibitem[Elliott and Burke(2009)Elliott, and Burke]{elliot09}
Elliott,~P.; Burke,~K. Non-empirical derivation of the parameter in the B88
  exchange functional. \emph{Can. J. Chem.} \textbf{2009}, \emph{87},
  1485\relax
\mciteBstWouldAddEndPuncttrue
\mciteSetBstMidEndSepPunct{\mcitedefaultmidpunct}
{\mcitedefaultendpunct}{\mcitedefaultseppunct}\relax
\EndOfBibitem
\bibitem[Elliott et~al.(2008)Elliott, Lee, Cangi, and
  Burke]{PhysRevLett.100.256406}
Elliott,~P.; Lee,~D.; Cangi,~A.; Burke,~K. Semiclassical origins of density
  functionals. \emph{Phys. Rev. Lett.} \textbf{2008}, \emph{100}, 256406\relax
\mciteBstWouldAddEndPuncttrue
\mciteSetBstMidEndSepPunct{\mcitedefaultmidpunct}
{\mcitedefaultendpunct}{\mcitedefaultseppunct}\relax
\EndOfBibitem
\bibitem[Chan and Handy(1999)Chan, and Handy]{PhysRevA.59.2670}
Chan,~G. K.-L.; Handy,~N.~C. Kinetic-energy systems, density scaling, and
  homogeneity relations in density-functional theory. \emph{Phys. Rev. A}
  \textbf{1999}, \emph{59}, 2670--2679\relax
\mciteBstWouldAddEndPuncttrue
\mciteSetBstMidEndSepPunct{\mcitedefaultmidpunct}
{\mcitedefaultendpunct}{\mcitedefaultseppunct}\relax
\EndOfBibitem
\bibitem[Nagy(2005)]{nagy:044105}
Nagy,~A. Density scaling and exchange-correlation energy. \emph{J. Chem. Phys.}
  \textbf{2005}, \emph{123}, 044105\relax
\mciteBstWouldAddEndPuncttrue
\mciteSetBstMidEndSepPunct{\mcitedefaultmidpunct}
{\mcitedefaultendpunct}{\mcitedefaultseppunct}\relax
\EndOfBibitem
\bibitem[Cohen et~al.(2012)Cohen, Mori-S\'{a}nchez, and
  Yang]{doi:10.1021/cr200107z}
Cohen,~A.~J.; Mori-S\'{a}nchez,~P.; Yang,~W. Challenges for density functional
  theory. \emph{Chem. Rev.} \textbf{2012}, \emph{112}, 289--320\relax
\mciteBstWouldAddEndPuncttrue
\mciteSetBstMidEndSepPunct{\mcitedefaultmidpunct}
{\mcitedefaultendpunct}{\mcitedefaultseppunct}\relax
\EndOfBibitem
\bibitem[Liu and Parr(1997)Liu, and Parr]{PhysRevA.55.1792}
Liu,~S.; Parr,~R.~G. Expansions of density functionals in terms of homogeneous
  functionals: Justification and nonlocal representation of the kinetic energy,
  exchange energy,and classical Coulomb repulsion energy for atoms. \emph{Phys.
  Rev. A} \textbf{1997}, \emph{55}, 1792--1798\relax
\mciteBstWouldAddEndPuncttrue
\mciteSetBstMidEndSepPunct{\mcitedefaultmidpunct}
{\mcitedefaultendpunct}{\mcitedefaultseppunct}\relax
\EndOfBibitem
\bibitem[Parr and Liu(1997)Parr, and Liu]{Parr1997164}
Parr,~R.~G.; Liu,~S. Some functional relations in the density functional theory
  of finite interacting electronic systems. \emph{Chem. Phys. Lett.}
  \textbf{1997}, \emph{276}, 164 -- 166\relax
\mciteBstWouldAddEndPuncttrue
\mciteSetBstMidEndSepPunct{\mcitedefaultmidpunct}
{\mcitedefaultendpunct}{\mcitedefaultseppunct}\relax
\EndOfBibitem
\bibitem[Cohen et~al.(2008)Cohen, Mori-S\'{a}nchez, and Yang]{cohen:121104}
Cohen,~A.~J.; Mori-S\'{a}nchez,~P.; Yang,~W. Fractional spins and static
  correlation error in density functional theory. \emph{J. Chem. Phys.}
  \textbf{2008}, \emph{129}, 121104\relax
\mciteBstWouldAddEndPuncttrue
\mciteSetBstMidEndSepPunct{\mcitedefaultmidpunct}
{\mcitedefaultendpunct}{\mcitedefaultseppunct}\relax
\EndOfBibitem
\bibitem[Cohen et~al.(2008)Cohen, Mori-S\'{a}nchez, and Yang]{Cohen08082008}
Cohen,~A.~J.; Mori-S\'{a}nchez,~P.; Yang,~W. Insights into current limitations
  of density functional theory. \emph{Science} \textbf{2008}, \emph{321},
  792--794\relax
\mciteBstWouldAddEndPuncttrue
\mciteSetBstMidEndSepPunct{\mcitedefaultmidpunct}
{\mcitedefaultendpunct}{\mcitedefaultseppunct}\relax
\EndOfBibitem
\bibitem[Borgoo and Tozer(2013)Borgoo, and Tozer]{doi:10.1021/ct400129d}
Borgoo,~A.; Tozer,~D.~J. Density scaling of noninteracting kinetic energy
  functionals. \emph{J. Chem. Theory Comput.} \textbf{2013}, \emph{9},
  2250--2255\relax
\mciteBstWouldAddEndPuncttrue
\mciteSetBstMidEndSepPunct{\mcitedefaultmidpunct}
{\mcitedefaultendpunct}{\mcitedefaultseppunct}\relax
\EndOfBibitem
\bibitem[Kohn and Sham(1965)Kohn, and Sham]{KS}
Kohn,~W.; Sham,~L. Self-consistent equations including exchange and correlation
  effects. \emph{Phys. Rev.} \textbf{1965}, \emph{140}, A1133\relax
\mciteBstWouldAddEndPuncttrue
\mciteSetBstMidEndSepPunct{\mcitedefaultmidpunct}
{\mcitedefaultendpunct}{\mcitedefaultseppunct}\relax
\EndOfBibitem
\bibitem[Clementi and Roetti(1974)Clementi, and Roetti]{CR74}
Clementi,~E.; Roetti,~C. Roothaan--Hartree--Fock atomic wavefunctions: basis
  functions and their coefficients for ground and certain excited states of
  neutral and ionized Atoms, Z <= 54. \emph{Atomic Data Nucl. Data Tables}
  \textbf{1974}, \emph{14}, 177\relax
\mciteBstWouldAddEndPuncttrue
\mciteSetBstMidEndSepPunct{\mcitedefaultmidpunct}
{\mcitedefaultendpunct}{\mcitedefaultseppunct}\relax
\EndOfBibitem
\bibitem[Iyengar et~al.(2001)Iyengar, Ernzerhof, Maximoff, and
  Scuseria]{IEMS01}
Iyengar,~S.~S.; Ernzerhof,~M.; Maximoff,~S.~N.; Scuseria,~G.~E. Challenge of
  creating accurate and effective kinetic-energy functionals. \emph{Phys. Rev.
  A} \textbf{2001}, \emph{63}, 052508\relax
\mciteBstWouldAddEndPuncttrue
\mciteSetBstMidEndSepPunct{\mcitedefaultmidpunct}
{\mcitedefaultendpunct}{\mcitedefaultseppunct}\relax
\EndOfBibitem
\bibitem[Biegler-k\"{o}nig et~al.(1982)Biegler-k\"{o}nig, Bader, and
  Tang]{KBT82}
Biegler-k\"{o}nig,~F.~W.; Bader,~R. F.~W.; Tang,~T.-H. Calculation of the
  average properties of atoms in molecules. II. \emph{J. Comput. Chem.}
  \textbf{1982}, \emph{13}, 317--328\relax
\mciteBstWouldAddEndPuncttrue
\mciteSetBstMidEndSepPunct{\mcitedefaultmidpunct}
{\mcitedefaultendpunct}{\mcitedefaultseppunct}\relax
\EndOfBibitem
\bibitem[Becke(1988)]{BEC88}
Becke,~A.~D. Density-functional exchange-energy approximation with correct
  asymptotic behavior. \emph{Phys. Rev. A} \textbf{1988}, \emph{38}, 3098\relax
\mciteBstWouldAddEndPuncttrue
\mciteSetBstMidEndSepPunct{\mcitedefaultmidpunct}
{\mcitedefaultendpunct}{\mcitedefaultseppunct}\relax
\EndOfBibitem
\bibitem[Perdew and Wang(1992)Perdew, and Wang]{PW91}
Perdew,~J.~P.; Wang,~Y. Accurate and simple analytic representation of the
  electron-gas correlation energy. \emph{Phys. Rev. B} \textbf{1992},
  \emph{45}, 13244--13249\relax
\mciteBstWouldAddEndPuncttrue
\mciteSetBstMidEndSepPunct{\mcitedefaultmidpunct}
{\mcitedefaultendpunct}{\mcitedefaultseppunct}\relax
\EndOfBibitem
\bibitem[TUR()]{TURBOMOLE}
{TURBOMOLE V6.2, 2009}, a development of {University of Karlsruhe} and
  {Forschungszentrum Karlsruhe GmbH}, 1989-2007, {TURBOMOLE GmbH}, since 2007;
  available from {\tt http://www.turbomole.com}.\relax
\mciteBstWouldAddEndPunctfalse
\mciteSetBstMidEndSepPunct{\mcitedefaultmidpunct}
{}{\mcitedefaultseppunct}\relax
\EndOfBibitem
\bibitem[Perdew et~al.(1996)Perdew, Burke, and Ernzerhof]{pbe}
Perdew,~J.~P.; Burke,~K.; Ernzerhof,~M. Generalized gradient approximation made
  simple. \emph{Phys. Rev. Lett.} \textbf{1996}, \emph{77}, 3865\relax
\mciteBstWouldAddEndPuncttrue
\mciteSetBstMidEndSepPunct{\mcitedefaultmidpunct}
{\mcitedefaultendpunct}{\mcitedefaultseppunct}\relax
\EndOfBibitem
\bibitem[Weigend and Ahlrichs(2005)Weigend, and Ahlrichs]{def2tzvpp}
Weigend,~F.; Ahlrichs,~R. Balanced basis sets of split valence, triple zeta
  valence and quadruple zeta valence quality for H to Rn: design and assessment
  of accuracy. \emph{Phys. Chem. Chem. Phys.} \textbf{2005}, \emph{7},
  3297\relax
\mciteBstWouldAddEndPuncttrue
\mciteSetBstMidEndSepPunct{\mcitedefaultmidpunct}
{\mcitedefaultendpunct}{\mcitedefaultseppunct}\relax
\EndOfBibitem
\bibitem[Rappoport and Furche(2010)Rappoport, and Furche]{furchepol}
Rappoport,~D.; Furche,~F. Property-optimized Gaussian basis sets for molecular
  response calculations. \emph{J. Chem. Phys.} \textbf{2010}, \emph{133},
  134105\relax
\mciteBstWouldAddEndPuncttrue
\mciteSetBstMidEndSepPunct{\mcitedefaultmidpunct}
{\mcitedefaultendpunct}{\mcitedefaultseppunct}\relax
\EndOfBibitem
\bibitem[Zhao and Truhlar(2005)Zhao, and Truhlar]{truhlar05a}
Zhao,~Y.; Truhlar,~D.~G. Design of density functionals that are broadly
  accurate for thermochemistry, thermochemical kinetics, and nonbonded
  interactions. \emph{J. Phys. Chem. A} \textbf{2005}, \emph{109},
  5656--5667\relax
\mciteBstWouldAddEndPuncttrue
\mciteSetBstMidEndSepPunct{\mcitedefaultmidpunct}
{\mcitedefaultendpunct}{\mcitedefaultseppunct}\relax
\EndOfBibitem
\bibitem[Zhao and Truhlar(2005)Zhao, and Truhlar]{truhlar05nb}
Zhao,~Y.; Truhlar,~D.~G. Benchmark databases for nonbonded interactions and
  their use to test density functional theory. \emph{J. Chem. Theory Comput.}
  \textbf{2005}, \emph{1}, 415--432\relax
\mciteBstWouldAddEndPuncttrue
\mciteSetBstMidEndSepPunct{\mcitedefaultmidpunct}
{\mcitedefaultendpunct}{\mcitedefaultseppunct}\relax
\EndOfBibitem
\bibitem[Constantin et~al.(2011)Constantin, Fabiano, and {Della Sala}]{zeta}
Constantin,~L.~A.; Fabiano,~E.; {Della Sala},~F. Improving atomization energies
  of molecules and solids with a spin-dependent gradient correction from
  one-electron density analysis. \emph{Phys. Rev. B} \textbf{2011}, \emph{84},
  233103\relax
\mciteBstWouldAddEndPuncttrue
\mciteSetBstMidEndSepPunct{\mcitedefaultmidpunct}
{\mcitedefaultendpunct}{\mcitedefaultseppunct}\relax
\EndOfBibitem
\bibitem[Constantin et~al.(2012)Constantin, Fabiano, and {Della Sala}]{vzeta}
Constantin,~L.~A.; Fabiano,~E.; {Della Sala},~F. Spin-dependent gradient
  correction for more accurate atomization energies of molecules. \emph{J.
  Chem. Phys.} \textbf{2012}, \emph{137}, 194105\relax
\mciteBstWouldAddEndPuncttrue
\mciteSetBstMidEndSepPunct{\mcitedefaultmidpunct}
{\mcitedefaultendpunct}{\mcitedefaultseppunct}\relax
\EndOfBibitem
\bibitem[Taut(1993)]{Taut}
Taut,~M. Two electrons in an external oscillator potential: Particular analytic
  solutions of a Coulomb correlation problem. \emph{Phys. Rev. A}
  \textbf{1993}, \emph{48}, 3561\relax
\mciteBstWouldAddEndPuncttrue
\mciteSetBstMidEndSepPunct{\mcitedefaultmidpunct}
{\mcitedefaultendpunct}{\mcitedefaultseppunct}\relax
\EndOfBibitem
\bibitem[Constantin et~al.(2011)Constantin, Chiodo, Fabiano, Bodrenko, and
  {Della Sala}]{JS}
Constantin,~L.~A.; Chiodo,~L.; Fabiano,~E.; Bodrenko,~I.; {Della Sala},~F.
  Correlation energy functional from jellium surface analysis. \emph{Phys. Rev.
  B} \textbf{2011}, \emph{84}, 045126\relax
\mciteBstWouldAddEndPuncttrue
\mciteSetBstMidEndSepPunct{\mcitedefaultmidpunct}
{\mcitedefaultendpunct}{\mcitedefaultseppunct}\relax
\EndOfBibitem
\bibitem[Beyhan et~al.(2010)Beyhan, G\"{o}tz, Jacob, and Visscher]{beyhan10}
Beyhan,~S.~M.; G\"{o}tz,~A.~W.; Jacob,~C.~R.; Visscher,~L. The weak covalent
  bond in NgAuF (Ng = Ar, Kr, Xe): A challenge for subsystem density functional
  theory. \emph{J. Chem. Phys.} \textbf{2010}, \emph{132}, 044114\relax
\mciteBstWouldAddEndPuncttrue
\mciteSetBstMidEndSepPunct{\mcitedefaultmidpunct}
{\mcitedefaultendpunct}{\mcitedefaultseppunct}\relax
\EndOfBibitem
\bibitem[Jacob and Visscher(2008)Jacob, and Visscher]{jacob08prot}
Jacob,~C.~R.; Visscher,~L. A subsystem density-functional theory approach for
  the quantum chemical treatment of proteins. \emph{J. Chem. Phys.}
  \textbf{2008}, \emph{128}, 155102\relax
\mciteBstWouldAddEndPuncttrue
\mciteSetBstMidEndSepPunct{\mcitedefaultmidpunct}
{\mcitedefaultendpunct}{\mcitedefaultseppunct}\relax
\EndOfBibitem
\bibitem[Kiewisch et~al.(2008)Kiewisch, Eickerling, Reiher, and
  Neugebauer]{neugebauer08}
Kiewisch,~K.; Eickerling,~G.; Reiher,~M.; Neugebauer,~J. Topological analysis
  of electron densities from Kohn--Sham and subsystem density functional
  theory. \emph{J. Chem. Phys.} \textbf{2008}, \emph{128}, 044114\relax
\mciteBstWouldAddEndPuncttrue
\mciteSetBstMidEndSepPunct{\mcitedefaultmidpunct}
{\mcitedefaultendpunct}{\mcitedefaultseppunct}\relax
\EndOfBibitem
\bibitem[Govind et~al.(2009)Govind, Sushko, Hess, Valiev, and
  Kowalski]{Govind09cpl}
Govind,~N.; Sushko,~P.; Hess,~W.; Valiev,~M.; Kowalski,~K. Excitons in
  potassium bromide: A study using embedded time-dependent density functional
  theory and equation-of-motion coupled cluster methods. \emph{Chem. Phys.
  Lett.} \textbf{2009}, \emph{470}, 353 -- 357\relax
\mciteBstWouldAddEndPuncttrue
\mciteSetBstMidEndSepPunct{\mcitedefaultmidpunct}
{\mcitedefaultendpunct}{\mcitedefaultseppunct}\relax
\EndOfBibitem
\bibitem[Fux et~al.(2010)Fux, Jacob, Neugebauer, Visscher, and Reiher]{fux10}
Fux,~S.; Jacob,~C.~R.; Neugebauer,~J.; Visscher,~L.; Reiher,~M. Accurate
  frozen-density embedding potentials as a first step towards a subsystem
  description of covalent bonds. \emph{J. Chem. Phys.} \textbf{2010},
  \emph{132}, 164101\relax
\mciteBstWouldAddEndPuncttrue
\mciteSetBstMidEndSepPunct{\mcitedefaultmidpunct}
{\mcitedefaultendpunct}{\mcitedefaultseppunct}\relax
\EndOfBibitem
\bibitem[Yan et~al.(1997)Yan, Perdew, Korhonen, and Ziesche]{monov}
Yan,~Z.; Perdew,~J.~P.; Korhonen,~T.; Ziesche,~P. Numerical test of the
  sixth-order gradient expansion for the kinetic energy: Application to the
  monovacancy in jellium. \emph{Phys. Rev. A} \textbf{1997}, \emph{55},
  4601\relax
\mciteBstWouldAddEndPuncttrue
\mciteSetBstMidEndSepPunct{\mcitedefaultmidpunct}
{\mcitedefaultendpunct}{\mcitedefaultseppunct}\relax
\EndOfBibitem
\end{mcitethebibliography}
\providecommand*\mcitethebibliography{\thebibliography}
\csname @ifundefined\endcsname{endmcitethebibliography}
  {\let\endmcitethebibliography\endthebibliography}{}

\end{document}